\documentclass[aps,reprint,twocolumn]{revtex4}
\usepackage{graphicx}        
\usepackage{amssymb}

\def\a{\alpha}
\def\b{\beta}

\def\d{\delta}
\def\ve{\varepsilon}
\def\e{\varepsilon}
\def\r{{\bf r}}

\def\s{\sigma}

\def\R{{\bf R}}

\begin{document}


\bibliographystyle{apsrev}
     

\title{Ribosome recycling, diffusion, and mRNA loop formation in 
translational regulation}

\author{Tom Chou \\
Dept. of Biomathematics, UCLA, Los Angeles, CA 90095-1766}
 \date{\today}              

\vspace{-8mm}
\begin{abstract}

\hspace{1.2in} $\mbox{Biophysical Journal, Volume 85, 755-773, 2003}$

\vspace{3mm}

\noindent {\bf ABSTRACT} We explore and quantify the physical 
and biochemical mechanisms that may be
relevant in the regulation of translation.  After elongation and detachment from the
3' termination site of mRNA, parts of the ribosome machinery can diffuse back to the
initiation site, especially if it is held nearby, enhancing overall translation rates.
The elongation steps of the mRNA-bound ribosomes are modeled using exact and
asymptotic results of the totally asymmetric exclusion process (TASEP) \cite{DER97}.
Since the ribosome injection rates of the TASEP depend on the local concentrations at
the initiation site, a source of ribosomes emanating from the termination end can feed
back to the initiation site, leading to a self-consistent set of equations for the
steady-state ribosome throughput.  Additional mRNA binding factors can also promote
loop formation, or cyclization, bringing the initiation and termination sites into close
proximity. The probability distribution of the distance between the initiation and
termination sites is described using simple noninteracting polymer models.  We find
that the initiation, or initial ribosome adsorption binding required for maximal
throughput can vary dramatically depending on certain values of the bulk ribosome
concentration and diffusion constant.  If cooperative interactions among the 
loop-promoting proteins and the initiation/termination sites are considered, the throughput
can be further regulated in a nonmonotonic manner. Potential experiments to test the
hypothesized physical mechanisms are discussed.

\end{abstract}

\maketitle    

\vspace{3mm}

\noindent Keywords: polymers, asymmetric exclusion process, protein production


\vspace{3mm}

\noindent {\bf INTRODUCTION}


The rate of protein production needs to be constantly regulated for all life
processes.  Genetic expression, protein production, and post-translational
modification, as well as transport and activation, are all processes that can
regulate the amount of active protein/enzymes in a cell. Although much recent
research has focused on the biochemical steps regulating the switching of genes
and rates of transcription, translational control mechanisms, post-translational
processing, and macromolecular transport are also important. For example, during
embryogenesis, nuclear material is highly condensed, transcriptional regulation
is inactive, and translational control is important \cite{DEVELOPMENT,WICKENS}. In other
instances, transcriptional regulation is accompanied by long lag times,
particularly with long genes.  Translational regulation is also the only means by
which RNA viruses express themselves.
 
\begin{figure}[!h]
\begin{center}
\includegraphics[height=5.9in]{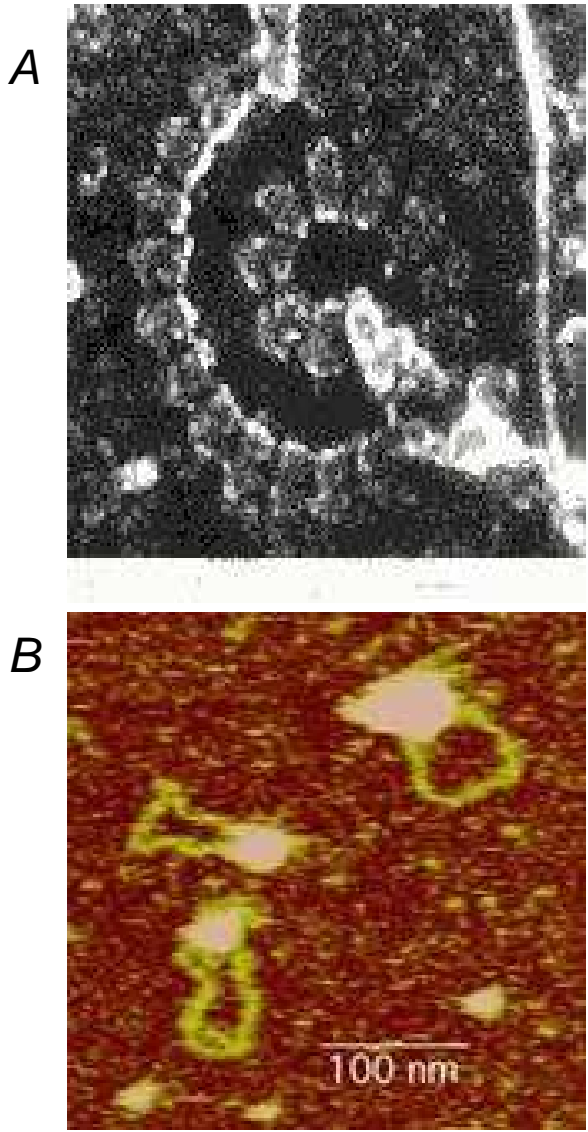}
\end{center}
\caption{$(A)$ An electron micrograph of 
polysomes on mRNA. 
$(B)$ An AFM micrograph of circularization of mRNA mediated by 
loop forming proteins. From Wells {\it et al.}, (1998). These images are of 
double stranded RNA of approximate length 2-4$\times$ the 
dsRNA persistence length. Single stranded end segments with 
loop binding factors comprise the ends.}
\label{AFM}
\end{figure}

Protein production, as with other cellular processes, requires the assembly of numerous
specific enzymes and cofactors for initiation. This assembly occurs in free solution and on
the 5' initiation site of mRNA. Translation involves unidirectional motion of the ribosome
complex along the mRNA strand as amino-acid-carrying tRNA successively transfer amino acids to
the growing polypeptide chain.  Images of mRNA caught in the act of translation often show
numerous ribosome complexes attached to the single-stranded nucleotide (Fig. \ref{AFM}$A$).
The multiple occupancy is presumably a consequence of very active translation, when many
copies of protein are desired.  

Under certain conditions, the local concentration of tRNA, ribosomes, initiation
factors, etc., will control protein production. One possible  {\it physical}
feedback mechanism underlying all the other biochemical regulation processes
utilizes local concentration variations of the components of translation
machinery.  Moreover, there is ample biochemical evidence that the 5' and 3' ends
of eukaryotic mRNA interact with each other, aided by proteins that bind to the
poly(A) tail and/or regions near the initiation site \cite{SAC90}, particularly
if the 5' initiation terminus is capped. The presence of both a poly(A) tail and
a 5' cap have been found to synergistically enhance translation rates in a number
of eukaryotic systems \cite{GAL91}.  Numerous proteins that initiate translation,
such as eukaryotic initiation factor eIF4, have been identified to bind to the
cap and initiate ribosomal binding \cite{SHIT,MUNROE,PRE99,SAC00}.  A different
set of proteins, poly(A) binding proteins (PAB) such as Pab1p, are found to bind
to the poly(A) tail. The proteins on the 5' cap and the poly(A) tail are also
known to form a complex (cap-eIF4E-eIF4G-Pab1p-poly(A) tail) which can increase
translation rates \cite{SCANNING,MUNROE,SAC97,SAC00}. {\it In vitro} solutions
of capped, poly(A)-tailed mRNA, tRNA, and ribosomes fail to display synergy
\cite{GAL91}, indicating that additional factors are required for cooperative
interactions between the cap and the poly(A) tail.  However, {\it in vitro}
systems that include caps, poly(A) tails, eIF's, and PAB's reveal circularized
mRNA structures in electron micrograph (EM) (Fig. 1$A$) and atomic force microscopy (AFM)
(Fig. 1$B$)
images.  In this way, it is thought that various components of the translation
machinery can be recycled after termination without completely reentering the
enzyme pool in the cytoplasm. 

Even in uncapped mRNA, there is evidence that certain sequences in the terminal 3'
untranslated region (UTR)  can enhance translation to levels comparable to those seen in
capped mRNAs \cite{MILLER97,SCANNING}.  Additionally, there are indications that proteins near
the termination end can, upon contact, directly activate \cite{GAL91} or inactivate
\cite{CURTIS,DUBNAU} ribosome entry at the 5' initiation site.  Loops also appear to be a
common motif in {\it DNA} structures \cite{LIBCHABER,HAGER96} and appear to take part in
transcriptional regulation \cite{HAGER96,DNALOOP,WYMAN}.  Double stranded DNA has a much
longer persistence length than single-stranded nucleic acids (such as mRNA) and is much less
likely to form loops without accompanying binding proteins or specific sequences. Direct
evidence for RNA ``circularization'' is shown in Figure \ref{AFM}$B$, which shows loop
formation of relatively short double-stranded mRNA in the presence of loop-binding factors at
their ends \cite{HAGER85}. It is reasonable to expect that the more flexible single-stranded
mRNA decorated with ribosomes can form similar loops.  Besides the AFM-imaged loop of double
stranded RNA shown in Fig. \ref{AFM}$B$, there is also substantial evidence, particularly in
viral mRNAs, that base pairing between uncapped 5' regions and non-polyadenylated 3' regions
forms closed loops of many kilobases \cite{MILLER97}.  This loop formation by direct base
pairing, or ``kissing,'' is a very plausible mechanism by which the 3' UTR recruits ribosomes
and delivers them to the 5' initiation site \cite{MILLER01}.

In this paper, we model the proposed cyclization, or ``circularization'' \cite{SAC97} and ribosome
recycling mechanisms.  Cooperative interactions of the initiation and termination sites with eIF's
and PAB proteins will also be considered within a number of reasonable assumptions.  Since
translation employs an immense diversity of mechanisms and proteins that vary greatly across
organisms \cite{SHIT}, we will only develop an initial, qualitative physical picture of cytoplasmic 
mRNA translation consistent with the ingredients mentioned above.  Three different coupled effects
are considered in turn: ($i$) a totally asymmetric exclusion process (TASEP) describing the
unidirectional stochastic motion of the ribosome along the mRNA, ($ii$) the diffusion and
adsorption/desorption kinetics from the mRNA initiation/termination sites, and ($iii$) the polymer
physics associated with how the termination and initiation sites are spatially distributed relative
to each other. The ribosome density along the mRNA, as well as the time-averaged throughput of
ribosomes, the ribosome ``current,'' are described by solutions of the TASEP.  The parameters in the
TASEP are the internal hopping rates and the injection and extraction rates at the initiation and
termination sites, respectively.  Since ribosome components that diffuse in bulk  must adsorb on the
initiation site, the injection rate used in the TASEP will be proportional to the local
concentration of the rate-limiting ribosome.  Ribosomes that reach  the termination site desorb and
reenter the pool of diffusing ribosomes.  The distance between the termination end and the
initiation site, when ribosomes are released, can thus influence the absorption rate and hence the
overall translation rate. The initiation-termination end-to-end distance distribution can be
estimated with basic polymer physics.  The end-to-end distance distribution can include effects such
as specific binding of poly(A) associated proteins with the 5' cap, thereby forming a loop, bringing
the initiation and termination sites into close proximity.  Although our model applies only to
cytoplasmic mRNA translation, many of its components can also be adapted to treat mRNA adsorption on
endoplasmic reticulum (ER) and ER-assisted translation.

\vspace{6mm}
\noindent {\bf PHYSICAL MODELS}
\vspace{3mm}

\begin{figure}[!h]
\begin{center}
\includegraphics[height=4.8in]{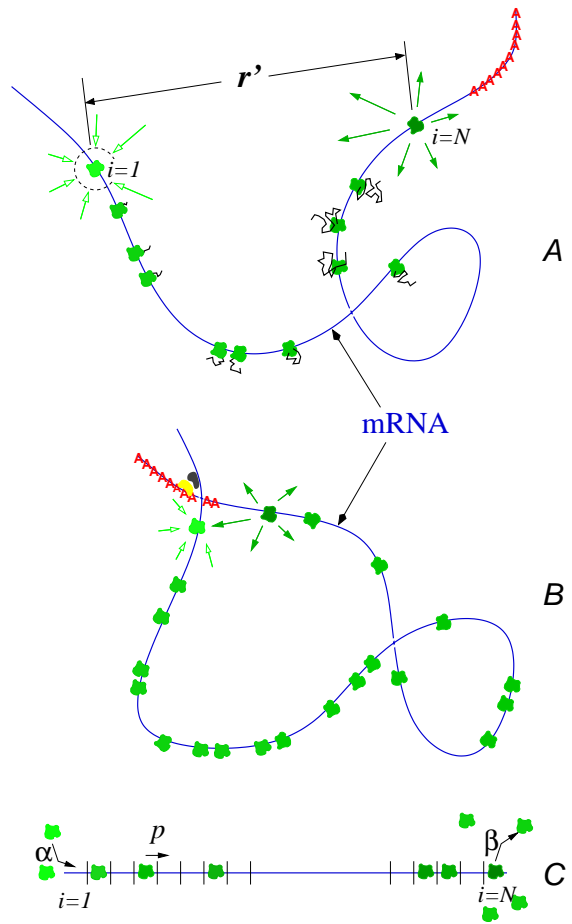}
\end{center}
\caption{A cartoon of mRNA translation in eukaryotes. The intermediary 
proteins and cofactors are not depicted. $(A)$ An mRNA chain loaded with
ribosomes (green), in various stages of protein (black) production.  
Ribosomal components as well as other components such as tRNA exist
at a uniform background concentration. The initiation and termination sites are
additional sinks ($i=1$) and sources ($i=N$), respectively, of ribosomes. 
$(B)$ Binding factors (yellow and dark grey) can increase the probability of loop
formation or ``circularization,'' which brings the poly(A) tail (red) in better
proximity to the initiation site, enhancing ribosome recycling.
$(C)$ Schematic of the associated TASEP 
with injection ($\a$), internal hop ($p$), and desorption ($\b$) 
rates labelled.}
\label{fig1}
\end{figure}

We now consider the physical processes necessary to describe the
above-mentioned translation processes.  At the relevant time scales, we will see
that fluctuations in these physical mechanisms are uncorrelated with each other.
This allows us to consider simple steady-states where time or ensemble averages
of the TASEP, ribosome diffusion in the cytoplasm, and the mRNA chain conformations are
uncorrelated and can be taken independently of each other.
A simplifying schematic of the
basic ingredients of mRNA translation is given in Fig. \ref{fig1}.

\vspace{2mm}
\noindent {\bf The Asymmetric Exclusion Process} 
\vspace{1mm}

The TASEP is one of a very small number of interacting nonequilibrium models with known exact
solutions. Asymmetric exclusion models have been used to effectively model qualitative features
of diverse phenomena including ion transport \cite{PORE1,CHOU99,CHOULOHSE}, traffic flow
\cite{TRAFFIC}, and the kinetics of biopolymerization \cite{BIOPOL1,BIOPOL2}.  Briefly, the model
consists of a 1D lattice of $N$ sites, each of approximately the
molecular size of a ribosome unit.  Each variable $\hat{\s}_{i} = \{0,1\}$ represents the
ribosome occupation at site $i$ of the coding region of mRNA.  Each site can be occupied by at
most one ribosome and the mean occupation $\s_{i} \equiv \langle\hat{\s}_{i}\rangle$ at each site
$1\geq \s_{i} \geq 0$.  The probability in time $\mbox{d}t$ that an individual ribosome moves
forward to the next site (toward the 3' end) is $p\mbox{d}t$, provided the adjacent site
immediately in front is unoccupied. Backward moves are not allowed, since ribosomes are strongly
driven motors that move unidirectionally from 5' to 3'. The entrance and exit rates at the
initiation $(i=1)$ and termination $(i=N)$ sites are denoted $\a$ and $\b$, respectively (cf. Fig
\ref{fig1}$C$).  The exact steady state solutions to this kinetic model, including the
average density $\s_{i}$, and the mean particle (ribosome) current have been found by Derrida and
Evans \cite{DER93}, using a matrix product ansatz, and by Sch\"{u}tz and Domany \cite{SCH93}, using
an iteration method.  An exact representation for the steady-state current across an $N$-site
chain is \cite{DER93}

\begin{equation}
J_{N}\equiv J(\a,\b,p) = p{S_{N-1}(p/\b) - S_{N-1}(p/\a) \over S_{N}(p/\b) - S_{N}(p/\a)},
\label{JEXACT}
\end{equation}

\noindent where 

\begin{equation}
\begin{array}{ll}
S_{N}(x) &  \displaystyle  = \sum_{k=0}^{N-1} {(N-k)(N+k-1)! \over N! k!}x^{N-k+1}
\end{array}
\label{SEXACT}
\end{equation}

In the $N\rightarrow \infty$ limit, the 1D TASEP (Eq. \ref{JEXACT}) admits three nonequilibrium
steady-state phases, representing different regimes of the steady state current
$J$:

\begin{equation}
\begin{array}{llll}

\mbox{(I)} & \displaystyle \a<{p\over 2}, \a<\b & \:\, \displaystyle J \equiv J_{L} =\a(1-{\a\over p}) & \:\, 
\s_{{N\over 2}} = {\a\over p} \\[13pt]
\mbox{(II)} & \displaystyle \b< {p\over 2}, \b<\a & \:\, \displaystyle J \equiv J_{R} =  \b(1-{\b\over p}) & \:\,
\s_{{N\over 2}} = 1-{\b\over p}  \\[13pt]
\mbox{(III)} & \displaystyle \a,\b\geq {p\over 2} &\:\, \displaystyle J\equiv J_{max}={p\over 4} & \:\,
\s_{{N\over 2}} = {1\over 2}.
\end{array}
\label{ASEMEQN}
\end{equation} 

\begin{figure}[t!]
\begin{center}
\includegraphics[height=3.0in]{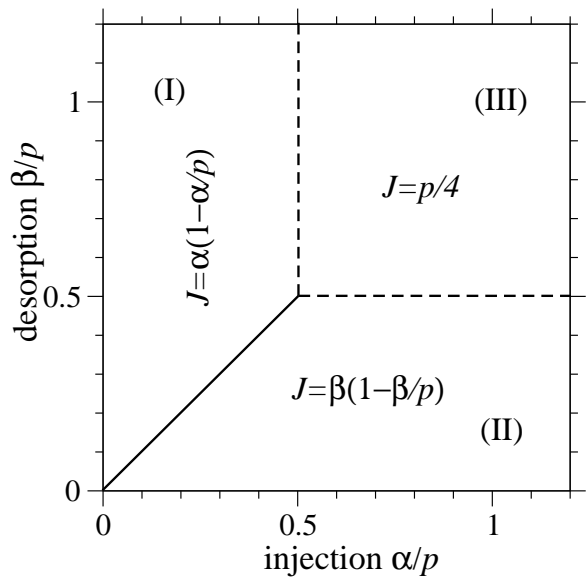}
\end{center}
\caption{The infinite chain ($N\rightarrow \infty$) limit 
nonequilibrium phase diagram of the standard TASEP. The maximal current (II), 
low density (I), and high density (II) phases and their corresponding 
steady state currents are indicated. In this and subsequent phase diagrams, solid curves 
correspond to phase boundaries across which the slope of the steady-state currents 
(with respect to the parameters) is discontinuous. Across the dashed phase boundaries,
the currents and their first derivatives are continuous.}
\label{PHASE0}
\end{figure}

The phases (I), (II), and (III) defined by Eqs. \ref{ASEMEQN} are denoted as the maximal current,
low density, and high density phases, respectively, and are delineated in Fig. \ref{PHASE0} by the
dotted phase boundaries.  Qualitatively, when $\b$ is small, and injection rates are faster than
extraction rates ($\a>\b$), the rate-limiting process is the exit step at $i=N$.  Therefore, the
high occupancy phase (II) has a low current which is a function of only the slow step $\b$.  In
the opposite limit of fast desorption at $i=N$, and slow injection at $i=1$ (small $\a$), the
chain is always nearly empty, and has a small current $J$ that depends only upon the rate
limiting step $\a$.  For large $\a \sim\b$, the system attains maximal current $J= p/4$ where the
effective rate-limiting steps are internal hopping rates $p$. In this phase, the constant current 
$J=p/4$ is independent of further increases in $\a$ or $\b$. The ribosomal currents given by
Eqs. \ref{ASEMEQN} and the associated phase diagram in Fig.  \ref{PHASE0} are valid only in the
$N\rightarrow \infty$ limit. Nonetheless, the $N=\infty$ phase diagram is qualitatively accurate
for the currents expected at large but finite $N$.  

There may appear to be a microphysical inaccuracy: The TASEP defined above corresponds to
individual movements with step length equal to the ribosome size.  However, ribosomes
typically occlude $\sim 10$ codons, so that it takes $\sim 10$ microscopic steps for the
ribosome to move the distance of its own size \cite{LAK03,SHAW03}. An accurate
approximation for the throughput $J$ (Eq. 1) is to assume that each step between two
sites defined in our model consists of $\sim 10$ actual tRNA transfers.  The effective
rate $p$ is thus the average tRNA transfer rate reduced by a factor of $\sim 10$. With
this consideration, the TASEP completely determines the steady state ribosome throughput
as long as the effective rate $p$ is appropriately defined. Therefore, we will treat the
mRNA translation problem using steps sizes equal to the ribosome size, with the
understanding that for appropriately rescaled transition rates, our results will be
qualitatively correct.  The exact currents of a TASEP, where the particle diameters are
$q$ times the step size, is given in Appendix B \cite{BIOPOL1}.  Explicit Monte-Carlo
simulations have also been performed on large particle$/$ small step size dynamics to
confirm the accuracy of the results \cite{LAK03,SHAW03}.

What remains is to determine the self-consistent dependence of the
model parameters, in particular $\a$ and $\b$, on the local ribosome concentration
(which in turn depends on the mean current $J$), diffusion rates, circularization,
etc. For example, the injection rate $\a$ at the initiation site will be proportional to
a microscopic binding rate  $k$ times the local ribosome concentration.

\vspace{2mm}
\noindent {\bf Steady-State Release, Diffusion, and Capture}
\vspace{1mm}

The complete mRNA translation machinery is extremely complicated, since it is  comprised
of many auxiliary RNA and protein cofactors, as well as a collection of active mRNA
chains.  Since there are many active mRNA chains in the cytoplasm, each mRNA chain feels
the sinks (initiation sites) and sources (termination sites) of all the other mRNA
chains. However, these other randomly distributed chains, each with their own initiation
and termination sites, contribute an averaged background ribosome concentration. Thus, it
is only the termination site (ribosome source) associated with the initiation site {\it
on the same} mRNA chain that resupplies the initiation site in a correlated manner.  We
thus consider a single ``isolated'' mRNA chain and for the sake of simplicity, assume
that a single component, say phosphorylated elongation initiation factor eIF4F or eIF2,
say \cite{CLEMENS,SAC00}, is key to a rate limiting step.  We will generically call this
component the ``ribosome.'' Consider a source of newly-detached ribosomes (emanating from
the 3' termination site) at position $\r$ away from the 5' initiation site.  The
probability of finding this particle within the volume element d$\r$ about $\r$ obeys the
linear diffusion equation with the termination site acting as a source,

\begin{equation}
\partial_{t}P(\r,t) - D\nabla^{2}P(\r,t) = J(t)W_{\it eff}(\r,t),
\label{DIFF}
\end{equation}

\noindent where $D$ is the bulk ribosome diffusion constant, $J(t)$ is the instantaneous rate of
ribosome release from the termination end, and $W_{\it eff}(\r)\mbox{d}\r$ is the probability
that the termination site is within the positions $\r$ and $\r + \mbox{d}\r$ from the initiation
site.  Although Eq. \ref{DIFF} can be solved exactly for all times, the TASEP result  (Eq. 
\ref{JEXACT}) is appropriate only in the steady-state, so we must consider that limit for all
processes. 

The typical mRNA passage time of a single ribosome is on the order of one minute.  The bulk
diffusion constant of the 10-15nm radius ($a\sim 15$nm) ribosome unit is $D
\sim 10^{-8}-10^{-7}$cm$^{2}$/s. A ribosome molecule will diffuse the length of a 1kB pair mRNA
strand in $\sim 0.1$s.  Therefore, with each release of a ribosome from the termination site, the
probability density appears as a pulse which passes through the initiation site over a time scale
shorter than it takes for a ribosome to stochastically hop a few lengths of its size along the
mRNA chain.  Therefore, an {\it upper} bound on the amount of correlation between concentration
fluctuations and $\hat{\s}_{1}$ can be found by considering the {\it equal} time two-point
correlation in the maximal current phase $\langle\hat{\s}_{1}\hat{\s}_{N}\rangle -\s_{1}\s_{N}
\sim N^{-3/2}/8$ \cite{CORR0}.  Two-point correlations in other current regimes are smaller, and
decay exponentially with $N$ \cite{CORR1}.  Therefore, we can neglect the correlation of the
current $J(t)$ with the occupancy $\hat{\s}_{1}$ at the initiation site.  Moreover, the
end-to-end distribution $W_{\it eff}$ arises from the statistics of the mRNA polymer configurations and
is also assumed independent of both $J(t)$ and $\hat{\s}_{1}$.
The steady state ribosome distribution can thus be found by setting
$\partial_{t}P(\r,t) = 0$ on the left-hand-side of Eq. \ref{DIFF} and taking the 
time, or ensemble, average of the remaining Poisson equation to obtain

\begin{equation}
\langle \nabla^{2}P(\r,t)\rangle  =  \nabla^{2}C(\r) = -{J\over D}W_{\it eff}(\r),
\label{LAPLACE}
\end{equation}

\noindent where $J\equiv \langle J(t)\rangle$ is the steady-state current of ribosomes 
emanating from the termination end of the mRNA re-entering the 
bulk ribosome pool, and $C(\r)=\langle P(\r,t)\rangle$ is the ensemble average of $P(\r)$.

The boundary condition for $C(\r)$ at the initiation site will depend on the occupancy of that
site.  When it is empty, there is a flux due to the microscopic adsorption step onto the first
site.  When $\hat{\s}_{1}=1$, the bulk ribosome probability distribution will obey perfectly reflecting
boundary conditions. Since the probability {\it at} $r=a$, $P(r=a,t)$ depends on the occupation
$\hat{\s}_{1}$, $\langle P(a,t) \hat{\s}_{1}\rangle \neq C(a) \s_{1}$. The mean concentration at
$r=a$ must be found by averaging the currents in the two states, $\hat{\s}_{1}=1$, and
$\hat{\s}_{1}=0$. When the initiation site is empty, 

\begin{equation}
J(\hat{\s}_{1}=0) \equiv J_{0} = 4\pi a^{2}D\partial_{r}C(r=a) = kC(r=a).
\label{BC1}
\end{equation}

\noindent Since the steady state current $J(\hat{\s}_{1}=1)=0$ when 
the initiation site is full, the 
averaged steady-state current is

\begin{equation}
J=(1-\s_{1})J(\hat{s}_{1}=0) +\s_{1}J(\hat{\s}_{1}=1) = (1-\s_{1})J_{0},
\label{BC1b}
\end{equation}

\noindent where $(1-\s_{1})$ is the fraction of time that the initiation site is unoccupied,
ready  to absorb a ribosome from the bulk.  This probability is not directly dependent on the
distribution $W_{\it eff}(\r)$, but will depend on the time-averaged local concentration $C(\r)$,
which in turn depends on $W_{\it eff}$ only through the distance of the source site at $i=N$.  

The solution to Eq. \ref{LAPLACE}, obeying the boundary conditions Eq. \ref{BC1} and
$C(r\rightarrow \infty) = C_{\infty}$, is 

\begin{equation}
\begin{array}{l}
\displaystyle C(\r) = C_{\infty} - {C_{\infty} \over r}\left({ka \over 4\pi Da 
+ k} \right) \\[13pt]
\:\hspace{2cm}\displaystyle  + {J\over D}\int d\r' G(\r-\r')W_{\it eff}(\r'),
\end{array}
\label{CR}
\end{equation}

\noindent where $\r$ is distance measured from the initiation site, and

\begin{equation}
\begin{array}{l}
\displaystyle G(\r,\r') = {1 \over 4\pi \vert \r-\r'\vert} - \\[13pt]
\:\hspace{1cm}\displaystyle \sum_{\ell=0,m=-\ell}^{\infty,m=+\ell}
\left[{ka^{\ell}-4\pi a^{2} D\ell a^{\ell-1} \over ka^{-\ell-1}+
4\pi a^{2} D(\ell+1)a^{-\ell-2}}\right] \\[13pt]
\: \displaystyle  \hspace{1.6cm} \times
{r_{<}^{-\ell-1}\over (2\ell+1)r_{>}^{\ell+1}}Y^{*}_{\ell m}(\Omega)Y_{\ell m}(\Omega')
\label{GREEN}
\end{array}
\end{equation}

\noindent is the associated Green function. In Eq. \ref{GREEN}, $r_{<}(r_{>})$ is the smaller(larger) of $\vert
\r\vert, \vert\r'\vert$ and $Y_{\ell m}(\Omega)$ are the spherical harmonic functions of the solid angle
$\Omega$ defined by the vector $\r$ \cite{ARF}. The first two terms in Eq. \ref{CR} arise from the uniform
concentration $C_{\infty}$ at infinity and the effects of a sink of radius $a$ at the initiation site.  The
sink decreases the effective concentration to a level below that of $C_{\infty}$. The last term proportional to
$J$ increases the local concentration and is the result of the source (termination site) some finite distance
away from the initiation site.  If $k\rightarrow 0$, and ribosomes do not bind even when the initiation site is
empty, the current $J$ must vanish, and $C(r) \rightarrow C_{\infty}$, as expected.  However, one cannot simply
consider the limit $k\rightarrow \infty$ in Eq.  (\ref{CR}) because $k$ and $\s_{1}$ are related through $J$,
the current determined by the TASEP in the rest of the chain. This can be seen by considering the limit
$k\rightarrow \infty$. If the rest of the TASEP contains the rate-limiting step to ribosome throughput, making
$J$ very small, it will effectively block clearance of the initiation site, since all sites of the chain will
be nearly occupied. In this case, $\s_{1} \approx 1$ and $k(1-\s_{1})$ is small (despite a large $k$), and
$C(r) \approx C_{\infty}$, as expected. However, if the rest of the chain is not rate-limiting, and if
clearance of the initiation site can occur fast enough, $\s_{1}< 1$ and $k(1-\s_{1})$ can be large. In this
case, $C(r)\approx C_{\infty}(1-a/r)+J D^{-1}\int d\r' G(\r-\r')W_{\it eff}(\r')$. The TASEP current $J$ will
eventually be balanced with $J=(1-\s_{1})J_{0}$. Note that $J$ is determined by Eq.  \ref{JEXACT} which in turn
{\it depends} on the entry rate $\a$ (in other words, $kC(a)$).  Thus, steady-state currents need to be {\it
self-consistently} determined, since $C(a)$ and $\s_{1}$ are not parameters, but dynamical variables that will
in turn be determined by setting $J=(1-\s_{1})J_{0}$. The analysis which uses Eq. \ref{JEXACT} to find
self-consistent explicit expressions for $J$ will be presented in the Results and Discussion.  

Since the averaged bulk concentration profile is spherically symmetric about the initiation site, only the
$\ell=0$ terms in the expression for $G(\r-\r')$ survive and 

\begin{equation}
\begin{array}{ll}
J_{0} & = \displaystyle kC(a) \\[13pt]
\: & \displaystyle = {4\pi a^{2} D kC_{\infty} \over ka+4\pi a^{2}D} 
+ {4\pi a^{2} D kJ\over 4\pi D(ka+4\pi a^{2} D)}\int_{r'>a} 
\!\!\!\!\!\!\!\! d\r' {W_{\it eff}(\r') \over r'} \\[13pt]
\: & \displaystyle = {k a \over k+4\pi a D}\left[4\pi DC_{\infty}+
{J\over R}\right],
\label{PHI}
\end{array}
\end{equation}

\noindent where 

\begin{equation}
{1\over R} \equiv \langle {1\over r}\rangle = \int \mbox{d}\r {W_{\it eff}(\r)\over r}.
\end{equation}

The surface concentration at the sink surface $a$ is reduced from the ``bulk'' value by a
factor of $1+4\pi a D/k$, due to adsorption and diffusional depletion.  However, part of
this initiation site concentration is also replenished at a rate proportional to the flux
$J$, due to the presence of a nearby termination (source) site.  The effects of this
replenishment are measured by the mean inverse separation $1/R$. The ``harmonic distance''
$R$ defines the effective distance felt by diffusing ribosomes as they make their way from
the termination end back to the initiation site. This particular $r^{-1}$ scaling is a
consequence of the solution to Poisson's equation (Eq. 5) in three dimensions, and is related
to the capture probability of diffusing ligands, as analyzed by Berg and Purcell \cite{BERG}.
Equation \ref{PHI} contains two unknowns, $C(a)$ and $\sigma_{1}$. We can use the
explicit solution Eq.  \ref{JEXACT} if we identify the injection rate $\a$ of the
TASEP with the unoccupied initiation site current $J_{0}=kC(a)\equiv \a$.  Equation
\ref{JEXACT} then relates $kC(a)$ to $\sigma_{1}$.  A second equation can be used by
noticing that the flux itself must be balanced.  Upon using $J=kC(a)(1-\s_{1})$ in Eq.
\ref{PHI}, a second relationship between $kC(a)$ and $\s_{1}$ can be found.
Substitution of the solution for $kC(a)$ (in terms of experimentally known or
controlled parameters $k,C_{\infty}, a, R, D$) into Eq. \ref{JEXACT} determines the
self-consistent, steady-state ribosome current.  This analysis, using the three
different explicit forms of Eq. \ref{JEXACT} (in the long chain limit) is presented in
the Results and Discussion.

\vspace{2mm}
\noindent {\bf End-to-End Distribution $W_{\it eff}$}
\vspace{1mm}

We now find  $W_{\it eff}(\r)$ in order to compute $R$ and obtain $C(a)$.  In some
cases, the mRNA chain may be anchored to cellular scaffolding or ER membranes such that
the initiation-termination separation is fixed. If one is interested in steady-state
protein production over a period which allows little change in initiation-termination
distance, $W_{\it eff}(\r) = \d(\r-\R)$, and $R = \vert\R\vert$.  In other cases, the
mRNA may be free to explore numerous conformations on the protein production time
scale.  Although it is possible that long mRNA strands may contain secondary structure,
we will assume that ribosomes, as they move along the mRNA, melt out these structures. 
Although there is evidence that mRNA can contain small, local loops \cite{HAGER85,MILLER97}, it
is less likely that they have larger-scale tertiary structure.  Thus, we will estimate
$W_{\it eff}$ and $R$ with simple polymer models.  

\begin{figure}[!h]
\begin{center}
\includegraphics[height=3.7in]{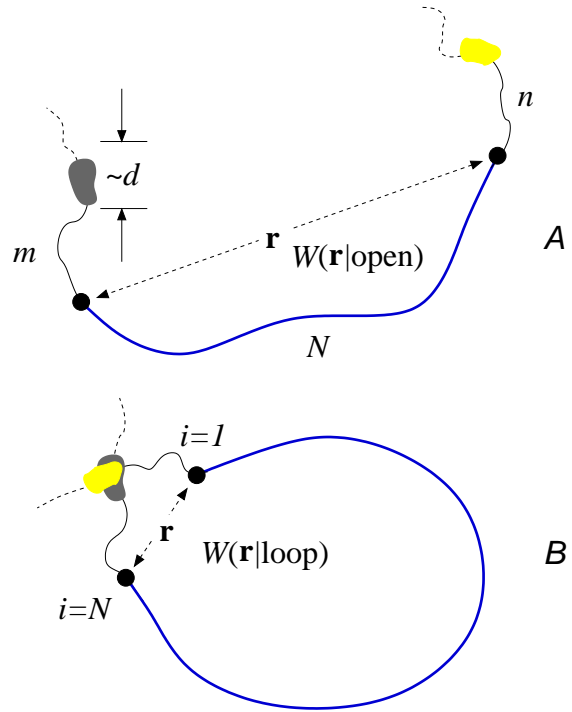}
\end{center}
\caption{A schematic of the effects of loop forming factors. 
The coding region of the mRNA is blue (the ribosomes and the poly-A tail
are not shown), the noncoding spacers of $m$ and $n$ persistence lengths
$\ve$ are solid black, while the the neglected short ends are dashed
curves. The loop binding factors are of typical size $d$.  $(A)$
Nonlooped conformations in which the initiation-termination site
distribution function is governed by $W(\r\vert \mbox{open})$.  $(B)$
The initiation-termination distribution function in looped
configurations is denoted $W(\r\vert \mbox{loop})$.  $W(\r\vert
\mbox{loop})$ is weighted more strongly at small $\vert\r\vert$ relative
to $W(\r\vert \mbox{open})$.  For stronger attraction between loop
binding factors the probability of loop formation increases, decreasing
the effective distance $R$ that ribosomes must diffuse to be recycled
back to the initiation site.}
\label{fig2}
\end{figure}

As shown in Fig. \ref{fig1}, the mRNA is comprised of three segments divided between
two qualitatively distinct regions. Typical coding regions are $\sim 10^{3}$ base
pairs, corresponding to $N\sim 300$. At low ribosome densities, the uncovered mRNA base
pairs will be rather flexible, and the effective persistence length $\ell$ will be a
local average between $a$ and the 2-4 nucleotide persistence length $\ve$ of uncovered
mRNA.  Large reductions in the persistence length of dsDNA containing segments of 
single-stranded regions have also been observed by Mills {\it et al.} \cite{SSPL}.
More sophisticated theories for variable persistence lengths can be
straightforwardly incorporated; however, for simplicity, we approximate the persistence
length in the coding region to be a uniform constant on the order of $\ell=a$, the
individual ribosome exclusion size.  The contour length of the coding region is thus
$L_{N}=Na$ with $N\sim 50-500$. The untranslated regions, or UTRs between the initiation
site and the binding factor (dark gray), and between the termination site and the loop-binding
factor (yellow), with persistence lengths $\ve$, have contour
lengths of $L_{m}=m\ve$ and $L_{n}=n\ve$, respectively.  Typical $L_{m}, L_{n}$ are on
the order of 100 bases so that $n, m \sim 20-50$.  However, extremely long
noncoding segments of order 1kbp can exist \cite{SHIT} where $m,n \sim 300$.  In what
follows we will also neglect all the excluded volume effects of the remaining short
ends of the mRNA chain.

As demonstrated by Wells {\it et al.} \cite{WEL98} in Figure \ref{AFM}$B$, mRNA can form
loops in the presence of binding proteins.  Therefore, we expect that $W_{\it eff}(\r)$ (and
hence $1/R$) will be a linear combination of $W(\r\vert\mbox{open})$ and
$W(\r\vert\mbox{loop})$, the initiation-termination probability distributions in open and
looped mRNA configurations, respectively.  These configurations are shown in Figs. 
\ref{fig2}$A,B$. For simplicity, we will use probability distributions associated with
noninteracting (phantom) chains and approximate the distributions $W(\r)$ with both a freely
jointed chain (FJC) and worm like chain (WLC) models with appropriate persistence lengths
$\ell$. The finite-sized, short distance behavior of the $W(\r\vert\mbox{open,\,loop})$ will
be important for accurately computing $\langle 1/r\rangle$.  As we will see,
$W(\r\vert\mbox{loop})$ can be constructed from the more fundamental quantity
$W(\r\vert\mbox{open})$ \cite{LIV95,SOKOLOV}.  Since we are eventually interested in either
ribosome transport from termination to initiation or in activation/deactivation of
initiation or release sites due to direct contact with the end proteins, we compute in
Appendix C the distance distribution $W(\r\vert\mbox{open})$ in the state where site $i=N$
is occupied and site $i=1$ is unoccupied.  

Using the $W(\r\vert\mbox{open})$ computed in Appendix C, we can thus consider the
contributions of looped configurations to the effective end-to-end distance
distribution. The binding energy between the 5'-cap and poly(A) tail proteins, $-U_{0}$
(in units of $k_{B}T$), determines the probability that the chain is looped:

\begin{equation}
\begin{array}{ll}
\displaystyle P_{loop}(n,m,N; U_{0}) & \displaystyle = {\exp(-G_{loop}) \over \exp(-G_{loop})
+\exp(-G_{open})} \\[13pt]
\: & =  \displaystyle {e^{U_{0}} \over e^{U_{0}}
+\Omega_{0}(\mbox{open})/\Omega_{0}(\mbox{loop})},
\end{array}
\end{equation}

\noindent where the free energies of a closed and open mRNA chain are $G_{loop} = -U_{0}-S_{loop}$
and $G_{open} = -S_{open}$, respectively.  Since the ratio of the number of configurations
under looped and open chain conditions is the ratio of probabilities of loop formation in the
{\it absence} of head-tail interactions ($U_{0}=0$), $\Omega_{0}(\mbox{open})/
\Omega_{0}(\mbox{loop}) = (1-P_{loop}^{(0)})/P_{loop}^{(0)}$, and

\begin{equation}
\displaystyle P_{loop} = {e^{U_{0}}P_{loop}^{(0)} 
\over e^{U_{0}}P_{loop}^{(0)}+(1-P_{loop}^{(0)})}.
\label{PLOOP}
\end{equation}

\noindent The probability that the ends of a noninteracting (in the absence of loop binding
proteins) chain intersects itself within the interaction volume defined by a thin spherical
shell of thickness $\delta$ (the binding interaction range) is approximately

\begin{equation}
\begin{array}{ll}
P^{(0)}_{\it loop} & \displaystyle \approx 4\pi d^{2}\delta 
\int_{r_{m}, r_{m+N}>d}\hspace{-13mm}
W_{\ve}(\r_{m}\vert\mbox{open})W_{a}(\r_{m+N}-\r_{m}\vert\mbox{open}) \\[13pt]
\: &\quad\times W_{\ve}(\r_{m+N+n}\vert\mbox{open})\mbox{d}\r_{1}\mbox{d}\r_{2} \\[14pt]
\: & \displaystyle \approx  \sqrt{{6\over \pi}}\left({d\over L_{T}}\right)^{2}\left({\delta \over 
L_{T}}\right)\left[1+O(d/L_{T})\right],
\end{array}
\label{PLOOP0}
\end{equation}

\noindent where $d$ is the typical size of the loop binding factors and $L_{T} \equiv
\sqrt{L_{N}^{2}+L_{m}^{2}+L_{n}^{2}} = \sqrt{Na^{2}+(m+n)\ve^{2}}$. We have assumed the total
radius of gyration $L_{T}\gg a$, and used a Gaussian chain as a qualitative approximation for the
distributions used in the calculation of  $P_{\it loop}^{(0)}$.
The conditional probability distribution  $W(\r\vert\mbox{loop})$ for
a looped chain is 

\begin{equation}
W(\r\vert\mbox{loop}) = {W_{a}(\r\vert\mbox{open})W_{\ve}(\r\vert\mbox{open})\over
\int_{r>a} W_{a}(\r'\vert\mbox{open})W_{\ve}(\r'\vert\mbox{open})\mbox{d}\r'},
\label{WLOOP}
\end{equation}

\noindent where $W_{\ell}(\r\vert\mbox{open})$ denotes the single segment, open chain
probability distributions in the two segments with persistence lengths $\ell = a, \ve$. For
$a\sqrt{N} \gg \ve\sqrt{m+n}$, the loop distribution given by Eq. \ref{WLOOP} is
qualitatively similar to the distribution function $W_{\ve}(\r\vert\mbox{open})$ of the
short segment of persistence length $\ve$.

\begin{figure}[!h]
\begin{center}
\includegraphics[height=5.2in]{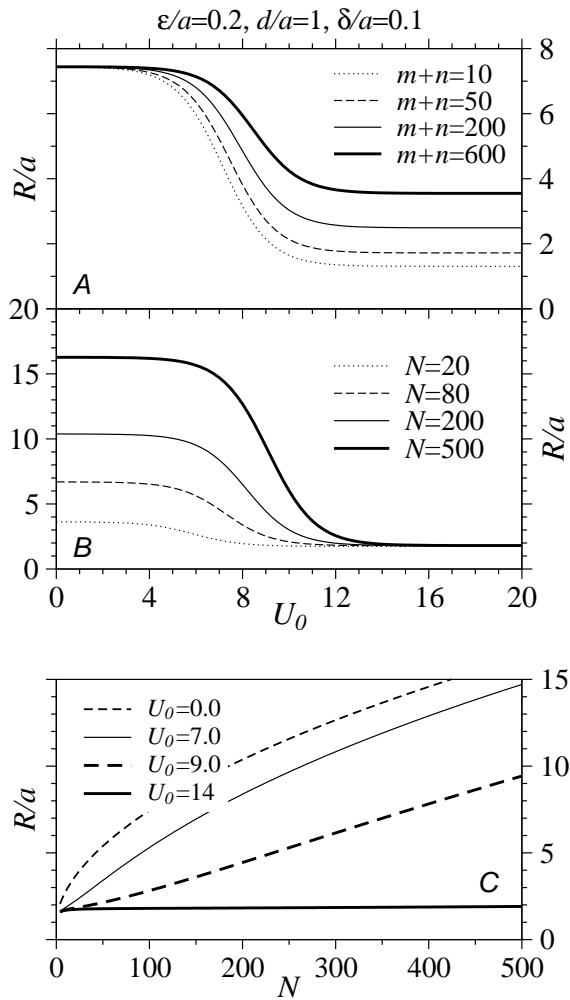}
\end{center}
\caption{The effective diffusional distance or ``harmonic
distance''  $R/a \equiv \left[a\int d\r W_{\it eff}(\r)/r\right]^{-1}$
over which recycled ribosomes must diffuse.  $(A)$ The dependence of
$R/a$ as a function of loop binding energy $U_{0}$ is shown for $N=100$
persistence lengths of coding mRNA. For large binding energies $U_{0}$,
the initiation and termination sites are brought closer together. The
crossover between the end-to-end distribution function of a free chain
to that of a loop occurs near $U_{0}^{*}\sim 8$. Increasing the length
of the short noncoding ends of the mRNA predominantly increases the
typical distance $R$ in the large $U_{0}$, looped regime.  $(B)$ The
$N$-dependence of $R/a$ with the ratio of noncoding persistence lengths
to coding persistence lengths $(m+n)/N = 1/2$. The $N$-dependence
manifests itself primarily in the low $U_{0}$, open chain regime. $(C)$
The $N$ dependence of $R/a$ for various $U_{0}$.}
\label{REFF}
\end{figure}

Using Eqs. \ref{PLOOP}, \ref{PLOOP0}, \ref{WLOOP}, and \ref{Wmn} we construct the 
effective initiation-termination distance distribution

\begin{equation}
W_{\it eff}(\r) = (1-P_{\it loop})W(\r\vert\mbox{open}) + P_{\it loop}W(\r\vert\mbox{loop}).
\label{WEFF}
\end{equation}

\noindent $W_{\it eff}(r)$ is plotted in Appendix C (Fig. \ref{PROB}) for various
$U_{0}$.  Qualitatively similar loop probability distributions have also been computed
by Liverpool and Edwards \cite{LIV95} within the WLC model but without finite-sized
molecules at the ends. Here and in all subsequent analyses, we use the typical
parameters $\ve/a =0.2, d=a$, and $\delta/a = 0.1$.  As $U_{0}$ is increased, the
distance distribution function switches over from $W(\r\vert \mbox{open})$ to
$W(\r\vert\mbox{loop})$.  The statistics of $W(\r\vert \mbox{open})$ and
$W(\r\vert\mbox{loop})$ are governed by $L_{N} = Na$ and $L_{mn} = (m+n)\ve$,
respectively.  The loop forming factors, since they are close to the initiation and
termination sites ($L_{mn}\ll L_{N}$), enhance the probability that the ends are close
to each other, particularly when the binding energy $U_{0}$ is large.  

The harmonic distance, $R$, determined using $W_{\it eff}$ is shown in
Figs.  \ref{REFF}$A,B$ as functions of loop binding energy $U_{0}$. The
result given by the last line in Eq. \ref{PLOOP0}, when used in Eqs.
\ref{PLOOP} and \ref{WEFF} qualitatively describes a crossover in
$W_{\it eff}$ from $W(\r\vert\mbox{open})$ to $W(\r\vert\mbox{loop})$
behavior at 

\begin{equation}
U_{0}^{*}\approx \ln\left[\sqrt{{\pi\over 6}}\left({L_{T}\over d}\right)^{2}{L_{T}\over \delta}\right]
+ O(d^{2}\delta/L_{T}^{3}). 
\end{equation}

In Fig. \ref{REFF}$A$, $R/a$ is shown with $N=100$, but at various noncoding
lengths $m+n$.  In the large binding strength limit, $R/a$ depends only on the
short distance $(m+n)\ve$. When loops rarely form, the typical separation between
initiation and termination sites can only depend on $L_{N}$ which is the only
quantity varied in Fig. \ref{REFF}$B$. Notice that the exact FJC solution (Appendix
C), or truncated WLC solution for $W(r\leq a\vert\mbox{open})=0$ ensures that
$R/a > 1$ for all values of $m,n,N$, and $U_{0}$.  The dependence of $R/a$ on $N$
is shown in Fig.  \ref{REFF}$C$ for various $U_{0}$. When $U_{0}$ is small, the
initiation-termination harmonic distance $R$ is controlled by $L_{N}$ and
increases as $\sqrt{N}$.  For larger $U_{0}$, the chain is partially bound into a
loop where the distance is controlled by the much shorter $L_{m+n}$.  The
harmonic distance $R$ remains small unless $N$ becomes extremely large so that
entropy can dominate and the loop ends can unbind.

We now couple our mathematical models by incorporating the $W_{\it eff}$-weighted
inverse harmonic distance $a/R$ into the local, effective concentration $C(a;R)$
given by Eq. \ref{PHI}. The effective injection rates $\a=kC(a)$ that control the translation
rate within  the steady-state TASEP are then self-consistently determined.

\vspace{6mm}
\noindent {\bf RESULTS AND DISCUSSION}
\vspace{2mm}

Here, we compute the possible currents $J$ and the parameter space in which each
are valid.  We will use the exact solution 
Eq. \ref{JEXACT}, or its three asymptotic forms (Eqs. \ref{ASEMEQN}), as well
as $J=kC(a)(1-\s_{1})$ in Eq. \ref{PHI}, to find all relevant quantities and
parameter phase boundaries.

%
%
%
%
%
%
%
%

Substitution of $J=kC(a)(1-\s_{1})$ into Eq. \ref{PHI} and solving for 
$\s_{1}$, we find

\begin{equation}
1-\s_{1} = {4\pi DR \over k}\left( 1- {C_{\infty}\over C(a)}\right) + {R \over a}.
\label{Y}
\end{equation}

\noindent Upon multiplying Eq. \ref{Y} by $kC(a)$, we find

\begin{equation}
\begin{array}{r}
\displaystyle kC(a)(1-\s_{1}) = 
4\pi D R (C(a)-C_{\infty}) + {R \over a}kC(a) \\[13pt] \equiv J(kC(a),\b,p)
= \left\{ \begin{array}{l}kC(a)(1-kC(a)/p) \\[11pt]
\b(1-\b/p) \\[11pt]
p/4 \end{array} \right.
\end{array}
\label{KC}
\end{equation}

\begin{figure}[!h]
\begin{center}
\includegraphics*[width=3.2in]{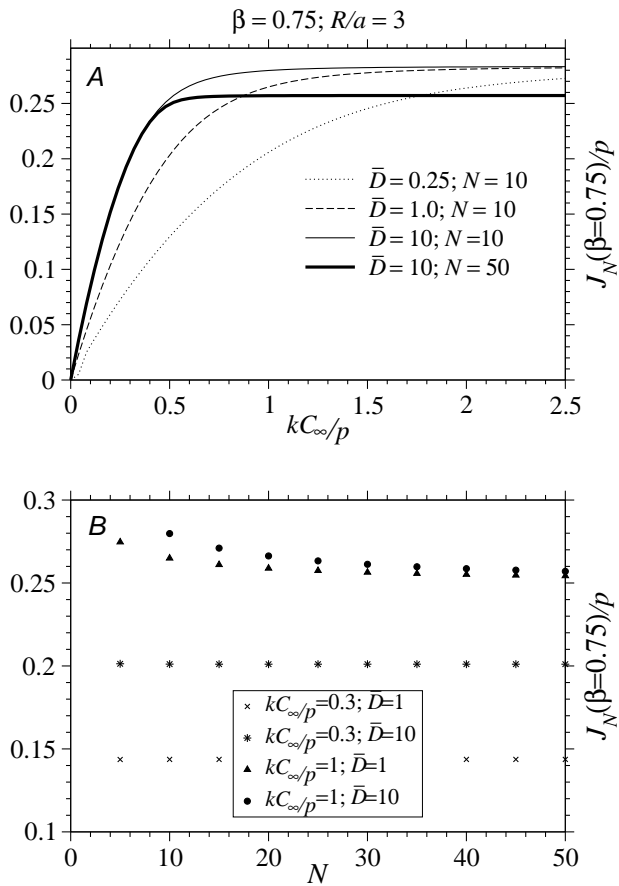}
\end{center}
\caption{The numerically determined, steady state currents at
finite $N$. The self-consistent currents were found by numerically
finding the roots to the polynomial in $J$ obtained by substituting the
last line of expression 19 into the exact equation 1.  $(A)$
Steady-state currents as a function of the injection rate
$kC_{\infty}/p$ for $R/a=3$ and various $\bar{D}=4\pi a
D/k=0.25,1.0,10$.  For $\bar{D}=10$, $N=10$ and $N=50$ are compared. 
$(B)$ $J$ as a function of length $N$ for $\bar{D}=1,10$ and
$kC_{\infty}/p=0.3,1$.}
\label{JEXACTFIG}
\end{figure}

\noindent To find $C(a)$ in terms of known parameters, we use the explicit solutions of the TASEP for the
current $J(kC(a),\b,p)$ (Eq. \ref{JEXACT} or \ref{ASEMEQN}) as indicated on the right-hand-side of Eq. 
\ref{KC}. The exact solution Eq.  \ref{JEXACT} yields an $N+2$ order equation in $kC(a)$ which we solve
numerically.  Only one of the $N+2$ roots of Eq. \ref{KC} is real, yields occupations between zero and one, and
is the physically relevant. The self-consistent solutions for $kC(a)$ are used to evaluate $J(kC(a),\b,p)$,
which are plotted in Figures \ref{JEXACTFIG}$A,B$.  As expected, shorter chains yield slightly higher current. 
Larger $D$ also increases the current and makes the approximate maximal current phase obtainable at smaller
$kC_{\infty}/p$.  Asymptotic limits for the current near phase boundaries and at large $N$ are given in
Appendix D.

The numerical solutions depicted in Fig. \ref{JEXACT} show, that 
for even modest $N\gtrsim 10$, the currents are accurately 
described by  their asymptotic expressions Eq. \ref{ASEMEQN}. 
Therefore, we can very accurately solve for $kC(a)$ and 
steady-state ribosome currents by separately considering 
each phase and its associated asymptotic form 
of $J$.

First assume that the detachment rate $\b \geq p/2$ and consider the maximal current (phase III in the TASEP)
where $J=p/4$. This occurs when both $\a,\b > p/2$.  To determine the parameter regime in which $J=p/4$ holds,
we solve for $C(a)$ and determine for what range of parameters $\a = kC(a) > p/2$. Using $J=p/4$ in Eq.
\ref{KC}, we find

\begin{equation}
C(a) = {p/4 + 4\pi DRC_{\infty} \over 4\pi D R + Rk/a}.
\label{KCMAX}
\end{equation}

\noindent The criterion for maximal current, $k> p/(2C(a))$, is thus

\begin{equation}
k > {p (4\pi D R + k(R/a)) \over p/2+8\pi D R C_{\infty}}.
\label{K1}
\end{equation}

\noindent Upon solving Eq. \ref{K1} for $k$, we find the minimum 
$k=k^{*}$ required to achieve maximal current $J=p/4$:

\begin{equation}
{k C_{\infty}\over p}>{k^{*}C_{\infty}\over p} = {1 \over 2 - {p \over 4\pi a DC_{\infty}}\left(1-{a\over 2R}\right)}.
\label{KSTAR}
\end{equation}

\noindent Note that for large enough $p/(4\pi a D C_{\infty})$ the critical 
value $k^{*}$ can diverge. The divergence is more likely or larger $R$
and occurs when there is simply not enough 
ribosome nearby to provide a large enough ``on'' rate $\alpha$ to 
achieve maximal current. Even when the source (termination end) is held at 
the initiation site ($R=a$), there is the possibility that 
$k^{*}$, and maximal current, are never attained. This 
behavior arises because even for ribosomes released at an 
infinitely absorbing spherical initiation surface, there 
is a probability of escape \cite{BERG}.

Next, let us consider small $\beta$ and large $\a=kC(a)$. The mRNA has a high
ribosome occupancy and a steady-state current $J = \beta(1-\beta/p)$. This regime
(phase II) is termination rate-limited and occurs for $\b<p/2$ and $\b < \a =
kC(a)$.  Upon using $J=\b(1-\b/p)$ in Eq. \ref{KC},

\begin{equation}
\b < kC(a) = k{\b(1-\b/p) + 4\pi DRC_{\infty} \over 4\pi DR + kR/a}.
\label{BETAKC}
\end{equation}

\begin{figure}[!h]
\begin{center}
\includegraphics*[width=3.3in]{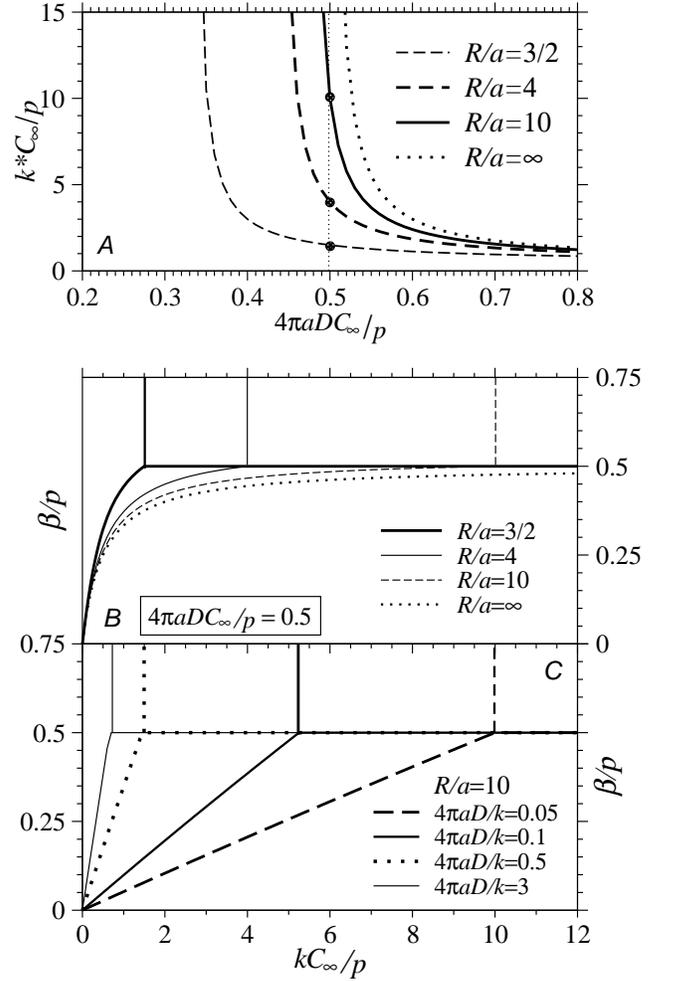}
\end{center}
\caption{The modified phase diagram for translation rates
along long $(N\rightarrow \infty$) mRNAs. $(A)$ The minimum binding rate
(Eq. \ref{KSTAR}) required to support the maximal current phase assuming
that $\beta > 1/2$.  This value depends on the bulk ribosome
concentration $C_{\infty}$ and the distance $R$ between the initiation
and termination sites. $(B)$ The modified phase diagrams as functions of
$kC_{\infty}/p$ for $4\pi a DC_{\infty}/p=0.5$ and various $R/a$.  $(C)$
Modified phase diagrams as functions of $kC_{\infty}/p$ for fixed
$\bar{D}=4\pi a D/k$ and $R/a=10$.}
\label{PHASE1}
\end{figure}

\noindent The only physical range of $\b$ that 
satisfies inequality \ref{BETAKC} is

\begin{equation}
\begin{array}{l}
\displaystyle \beta < \beta^{*}(k) = 
{p\over 2}\left({R\over a}(\bar{D}+1)-1\right) \\[13pt]
\: \displaystyle \hspace{1cm}\times \left[\sqrt{1+ {4(R/a)\bar{D}kC_{\infty} \over 
p((\bar{D}+1)R/a-1)^{2}}}-1\right],
\end{array}
\label{BSTAR}
\end{equation}

\noindent where $\bar{D} \equiv 4\pi a D/k$.  Equation \ref{BSTAR} defines the phase boundary between the
high-density, exit rate-limited phase (II) and the low-density, initiation rate-limited phase (I). This phase
boundary is plotted as a function of $kC_{\infty}/p$ for fixed $4\pi a D C_{\infty}/p = 0.5$ in Figs.
\ref{PHASE1}$B$. In the limit $kC_{\infty}/p \rightarrow 0$, the phase boundary straightens as in the standard
TASEP and is approximately

\begin{equation}
{\beta^{*}\over p} = {kC_{\infty} \over p}\left[1 - {(1-a/R) k \over  4\pi a DR} + 
O(k^{2})\right].
\end{equation}

Finally, when $\b > \b^{*}(k)$, but the entrance rate $kC(a)$ is low ($<p/2$), a low density phase
with $J = \a(1-\a/p)= kC(a)(1-kC(a)/p)$ exists.  The phase boundary delineating the low density
phase (I) is defined by $k< k^{*}$ and $\beta = \b^{*}(k)$. Upon using the current
$J=kC(a)(1-kC(a)/p)$ in Eq. \ref{KC}, we find $kC(a) = \beta^{*}$, and the current 
in the initiation rate-limited 
phase (I):

\begin{equation}
\begin{array}{l}
\displaystyle  J_{L} = {p\over 2}{R\over a}(\bar{D}+1)
\left({R\over a}(\bar{D}+1) -1\right)\  \\[13pt]
\:\displaystyle  \hspace{1cm}\times \left[\sqrt{1+{4(R/a)\bar{D}kC_{\infty}\over
p((\bar{D}+1)R/a-1)^{2}}}-1\right]-4\pi DR C_{\infty}.
\end{array}
\label{JEXACTB}
\end{equation}

\noindent In the limit $p/(kC_{\infty})\rightarrow \infty$, 

\begin{equation}
\begin{array}{l}
\displaystyle J_{L}(p\rightarrow \infty) = 
{4\pi a D k C_{\infty} \over 4\pi a D + k(1-a/R)} - \\[13pt]
\: \quad \displaystyle {(4\pi a D)^{2}(k+4\pi a D)k C_{\infty} \over 
(k(1-a/R) + 4\pi a D)^{3}}\left({kC_{\infty} \over p}\right) + O(p^{-2}),
\end{array}
\label{JLASYMP}
\end{equation}

\noindent which reduces to the result one would expect from infinitely fast 
initiation site clearance.

Summarizing, the large-$N$ steady-state ribosome currents 
(given by Eq. \ref{JEXACT}) in terms of ribosome concentrations and 
kinetic ``on'' rates are 

\begin{equation}
\begin{array}{lll}
\mbox{(I)} &\displaystyle  k<k^{*}, \b>kC(a) & \:\, \displaystyle J \equiv J_{L} =kC(a)(1-kC(a)/p) 
\\[13pt]
\mbox{(II)} & \displaystyle \b< {p\over 2}, \b<kC(a) & \:\, \displaystyle J \equiv J_{R} =  
\b(1-{\b\over p}) \\[13pt]
\mbox{(III)} &\displaystyle  k>k^{*},\b\geq {p\over 2} &\:\, \displaystyle J\equiv J_{max}={p\over 4}.
\end{array}
\label{ASEMEQN2}
\end{equation} 

\noindent where $kC(a)$ in phase (I) is expressed in terms of known parameters according
to Eq. \ref{KC}. The mean occupations of the initiation and termination sites, in each
regime, can now be readily found. At the first site, $\s_{1} = 1-J/(kC(a))$, where we use
$J=J_{L}, \b(1-\b/p)$, or $p/4$ (currents associated with each phase), and $kC(a)$ found
from Eqs. \ref{KC}, \ref{BETAKC}, or \ref{KCMAX}.  Similarly, the occupation at the last
site is $\s_{N} = J/\b$. All of our results can be expressed in terms of three of the four
nondimensional parameters: $\bar{D}=4\pi a D/k$, $kC_{\infty}/p$, $4\pi a
DC_{\infty}/p$, and $R/a$. We shall present our results in terms of the relevant
nondimensional parameters appropriate for the discussion at hand.

Figure \ref{PHASE1}$A$ shows the critical value $k^{*}$, above which an $N\rightarrow
\infty$ TASEP is in the maximal current phase (provided $\beta/p > 1/2$).  When
$C_{\infty}$ is small and $p$ is large, there is not enough ribosome in the cytoplasm to
feed the initiation fast enough compared to the clearance rate $p$.  Therefore the
maximal current ($J=p/4$) arises only when the binding is efficient and $k>k^{*}$ is
large.  For smaller $R$ (termination site close to the initiation site), smaller values
of $4\pi aDC_{\infty}/p$ can still support maximal current.  From Eq. \ref{KSTAR}, we see
that when $4\pi aDC_{\infty}/p \leq (1-a/(2R))/2$, the critical value $k^{*}$ diverges
and the maximal current can never be reached.  There is simply not enough ribosomes or
the diffusion is too slow for there to be sufficient concentration at the initiation site
to support the maximal current phase.

If the diffusion constant $D$ and $C_{\infty}$ are chosen such that, for example, $4\pi a
DC_{\infty}/p$ is small, the critical values $k^{*}$ vary considerably with $R/a$, as shown by the green
points ($4\pi a D C_{\infty}/p = 1/2$) in Fig. \ref{PHASE1}$A$.  The effects of depletion arise suddenly, with onset only at
values of $4\pi aDC_{\infty}/p \lesssim 0.6$.  For large $R/a$, values of $4\pi
aDC_{\infty}/p\sim 0.5$ will render the critical $k^{*}$ values very sensitive to $R$.  If the
initiation site has an interaction size of $a\sim 10$nm, and $p\sim 2-3$/s (20-30 codons/s)
\cite{CODON}, a diffusion constant of $D \sim 10^{-8}$ cm$^{2}$/s requires an effective
concentration of $C_{\infty}\sim 0.01-0.02\mu$M for the phase diagram to be sensitive to
diffusional depletion and $R$.  Although typical total cytoplasmic ribosome concentrations are
$C_{\infty} \sim 1\mu$M, many components must assemble in order to activate a translation-viable
ribosome.  For example, eIF4F exists at 0.01-0.2 times the total ribosome concentration
\cite{DUNCAN}. Furthermore, this already low abundance of eIF often needs to be further
phosphorylated to be active. Thus, the effective concentrations $C_{\infty}$ (and even diffusion
constants) appropriate for our model may very well be low enough to fall within the range for the
phase boundaries to be extremely sensitive to diffusional effects.

Figures \ref{PHASE1}$B,C$  show the steady-state phase diagrams as functions of $\beta/p$ and effective binding
rate $kC_{\infty}/p$. In these phase diagrams, as in the unperturbed ones defined by Eq. \ref{ASEMEQN}, the
upper left region corresponds to a low density phase, the lower right region corresponds to a high density
phase, and the upper right region describes a half-occupied (except near the ends $i=1,N$), maximal current
phase. The current $J$ is constant throughout the maximal current phase and is not changed if $kC_{\infty}/p$
or $\b$ is increased beyond $k^{*}C_{\infty}/p$ and $1/2$, respectively.  The phase diagram is modified by
ribosome diffusion and depletion near the initiation site.  The unmodified phase boundary between phases (I)
and (II) of the TASEP (Eq.  \ref{ASEMEQN}) would simply be defined by the straight line segment $\b/p =
kC_{\infty}/p$. The main effects of diffusional depletion (by the initiation sink) and replenishment (by the
termination source) on the standard phase diagram Fig.  \ref{PHASE0} is to shift the low density-maximal
current phase boundary to larger effective injection rates $kC_{\infty}/p$ and bend the low density-high
density phase boundaries accordingly.  Figure \ref{PHASE1}$B$ depicts the phase boundaries defined by Eqs.
\ref{KSTAR} and \ref{BSTAR} for fixed $R/a = 3/2,4,10,\infty$, and fixed $4\pi aD C_{\infty}/p = 1/2$ as
indicated by the green points in Fig. \ref{PHASE1}$A$. In this example $k^{*}C_{\infty}/p = 3/2, 4, 10$ for
$R/a = 3/2, 4, 10$, respectively. Note that for $R/a\rightarrow \infty$ that $k^{*}$ diverges and the maximal
current phase is never attained. If $4\pi aD C_{\infty}/p < 1/2$, then there will be a finite value of $R/a$
such that $k^{*}$ diverges.

If, instead,  $\bar{D} = 4\pi a D/k$ is held fixed, the phase boundaries are nearly
straight, as shown in Fig. \ref{PHASE1}$C$. Here, we fixed $R/a=10$, and plotted the
phase diagrams for $\bar{D} = 0.05, 0.1, 0.5, 3$. The corresponding values of
$kC_{\infty}/p$ above which the maximal current phase is attained  are $k^{*}C_{\infty}/p
= (1/2)(1+(1-a/(2R))/\bar{D} = 10, 21/4, 29/20,$ and $79/120$, respectively.

Our results up to this point are contingent on the fact that 
measurements are averaged over time scales such that the 
TASEP and the diffusion processes have reached steady-state, and the
mRNA chain distribution has thermally equilibrated. The possibility exists that 
the chain conformations are not in thermodynamic equilibrium 
while the TASEP and the bulk ribosome diffusion has reached steady-state
for a given chain conformation.
Thus, although not relevant within each of the three well-defined
physical processes, the issue of kinetic {\it versus} thermodynamic control 
of ribosome throughput arises when one considers measurements 
over time scales that are insufficient to allow equilibration of the mRNA 
chain. The consequences of this are discussed in the following section.

\vspace{6mm}
\noindent {\bf EXPERIMENTAL CONSEQUENCES AND PROPOSED MEASUREMENTS}
\vspace{2mm}

The basic physical mechanisms described in our model for mRNA translation suggest
a number of experimental tests. However, it must be emphasized that the model is
meant to provide qualitative guidelines most useful for studying trends and how
they depend on {\it physical} parameters.  Translation occurring {\it in
vivo} involve too many molecular species and 
biochemical processes to be quantitatively modeled, especially in the absence of significantly more detailed
experimental findings.  Nonetheless, our proposed mechanisms can be probed with
carefully designed, simplified, {\it in vitro} experiments.  Here, we discuss in
detail the basic expected phenomena and their regimes of validity.  

First note from Figure \ref{JEXACTFIG} and from Appendix D that the exact
currents for a finite number of codons $N$ very rapidly approach the asymptotic
values given by Eq. \ref{ASEMEQN} as $N$ increases. Even when $N$ is only $\sim
10-50$, the steady-state ribosome currents are only a few percent off the exact
$N=\infty$ results. In other words, the exact solution Eq. \ref{JEXACT} is a very
good approximation to Eq. \ref{ASEMEQN} for $N\gtrsim 10$. Therefore, as a mental
guide, it is typically sufficient to consider the currents $J$ corresponding to an
infinite chain ($N=\infty$) given by Eq. \ref{ASEMEQN}, but nonetheless consider
a finite initiation-termination separation (measured by the harmonic distance
$R$).  

\vspace{2mm}

\noindent {\bf Polysomal Density Variations} 
\vspace{1mm}

\noindent Although we have focused on the steady-state current, the particle
(ribosome) densities in each of the three current regimes are different and may be
detected. In the TASEP model, the ribosome density profiles along the mRNA chain vary
only near the initiation and termination ends. In the interior of the mRNA, the density
is relatively uniform and are given by the last column in Eqs.  \ref{ASEMEQN}. In the
exit-rate limited phase (small $\b/p$), where $J=\b(1-\b/p)$, the midpoint density
$\sigma_{N/2} \sim 1-\b/p$ is high, while in the low injection rate case,
$J=\a(1-\a/p)$, and $\s_{N/2} \sim \a/p$ is low. The typical density in the maximal
current regime is $\s \sim 1/2$. These densities are also approximately correct when
one explicitly treats large ribosomes that occlude many codon ``lattice sites.''
Therefore, we might expect that one may be able to predict in which current regime
translating mRNA exists if ribosome densities can be estimated from images taken with
{\it e.g.} AFM or EM techniques. For example, in Figure \ref{AFM}$A$, the high density of 
ribosomes suggests that the system is in phase (II) where the steady-state current $J = 
\b(1-\b/p)$ is a function only of the detachment rate $\b$.

\vspace{2mm}

\noindent {\bf Kinetic Binding Rate and Ribosome Concentration Dependences} 
\vspace{1mm}

\noindent Figure \ref{PHASE1}$C$ shows the minimum effective attachment rate 
$k^{*}C_{\infty}/p$
necessary for a large system to be in the maximal current regime (where the
ribosome current $J \approx p/4$) as a function of the effective ribosome
diffusion constant.  An additional requirement is that the effective detachment
rate $\b/p > 1/2$.  The value of $k^{*}$ can be tuned perhaps by substitution of
the codons comprising the initiation sites, or by other physical means. Although
ribosome diffusion constants are difficult to vary over a wide range (by
modifying the solution viscosity), the critical $k^{*}$ is a very sensitive
function of $D$, particularly for small $D$.  It is thus possible that slightly
increasing the ribosome diffusivity can dramatically decrease the $k^{*}$
necessary for the system to be in the maximal current regime.

As mentioned, changing the mRNA length $N$ does not significantly affect the overall
steady-state current along the chain (beyond about $N\sim 10-20$) but it {\it can}
change the statistics of the initiation-termination separation by changing $R$. 
Increasing the harmonic separation $R$ has qualitatively the same effect as decreasing
the ribosome diffusivity, since terminated ribosomes now have further to diffuse back to
the initiation site. For 

\begin{equation}
D < {p (1-a/(2R)) \over 8\pi a C_{\infty}},
\label{DCRIT}
\end{equation}

\noindent the maximal current regime is never reached.  This can be easily seen from equation
\ref{KSTAR}.  Thus, rather than tuning the ribosome diffusivity, decreasing $C_{\infty}$ may
preclude the system from entering the maximal current phase if Eq. \ref{DCRIT} is satisfied.
There is simply not enough ribosome available for sufficient initiation to be achieved so that
the maximal current phase arises.

\begin{figure}[!h]
\begin{center}
\includegraphics*[width=3.3in]{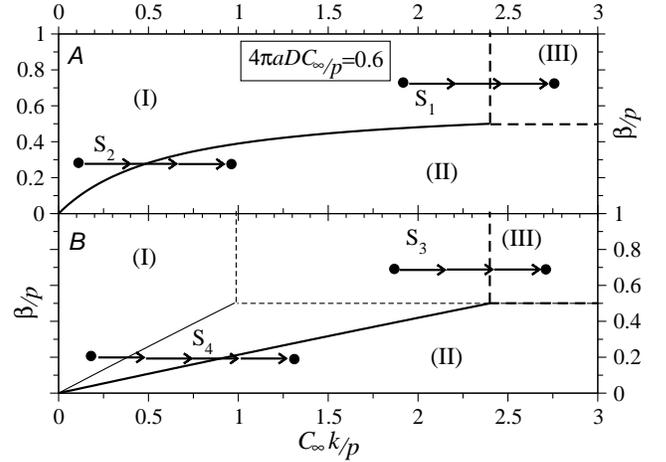}
\end{center}
\caption{Large $N$ phase diagrams for $R/a=10$. $(A)$ Phase
diagram for fixed $4\pi aDC_{\infty}/p = 0.6$ with trajectories
$S_{1,2}$ corresponding to increasing kinetic ``on'' rate $k$.  $(B)$
Phase diagram when $\bar{D}=4\pi aD/k =1,0.25$ is fixed, and
trajectories $S_{3,4}$ correspond to increasing bulk ribosome
concentration $C_{\infty}$. Trajectory $S_{3}$ traverses the (I)-(III)
phase boundary for $\bar{D}=0.25$ (thick curves) but not for
$\bar{D}=1.0$ (thin curves).  Trajectory $S_{4}$ on the other hand,
traverses the (I)-(II) phase boundaries for both $\bar{D}=0.25,1.0$.}
\label{BBAR_IDIOT}
\end{figure}

When inequality \ref{DCRIT} is not satisfied, the maximal current phase can exist.  In
Figure \ref{BBAR_IDIOT}$A$, we replot the phase diagram corresponding to $R/a=10$
shown in Fig. \ref{PHASE1}. Fixing the parameter $4\pi aDC_{\infty}/p=0.6$ allows $k$ to
be the only free parameter.  This kinetic ``on'' rate $k$ can be tuned by
varying ribosome recruitment proteins such as eIF4E.  If $\beta/p > 1/2$, $C_{\infty},
D,$ and $p$ are held constant, increasing $k$ from a sufficiently small value
allows one to traverse the trajectory $S_{1}$. The steady-state ribosome current
starts in the low density phase (I) with current given by Eq. \ref{JEXACTB}. As
$k$ is increased, the steady-state current increases until it continuously
crosses over into the maximal current regime (III), where the ribosome throughput is
given by $J=p/4$.  Further increasing $k$ when inside the maximal current phase
(III) will no longer affect the steady-state ribosome current. If, however, $\beta/p
< 1/2$, the current behavior abruptly crosses over (along trajectory $S_{2}$) from
that given by Eq. \ref{JEXACTB} to $J=\beta(1-\beta/p)$ corresponding to the high
ribosome density phase (II).  In this phase the detachment step is rate-limiting,
and further increases in $k$ will no longer affect the throughput.

If $k$ is held fixed and the ribosome concentration is independently varied instead, it is more
instructive to plot the phase diagram for fixed $\bar{D}\equiv 4\pi a D/k$ and $R/a$, as shown in
Fig.  \ref{BBAR_IDIOT}$B$.  Here, we choose the representative values $R/a=10$ and $\bar{D}=4\pi
aD/k=0.25,1$ and motivate parameter trajectories obtained by varying only $C_{\infty}$.  For
$\beta/p > 1/2$, increasing the bulk ribosome concentration traces out the trajectory $S_{3}$
continuously from the low density phase (I) (Eq. \ref{JEXACTB}) to the maximal current ($J=p/4$)
phase.  Further increasing the concentration well into  the maximal current phase will no longer
affect the throughput. Similarly, if $\beta/p< 1/2$, increasing $C_{\infty}$  can shift the
behavior from that of the low density phase to that of the high density, exit rate-limited phase.
Alternatively, one may vary $p$, the mean elongation rate of individual ribosomes, by controlling
the tRNA concentration in solution.  For example, decreasing available tRNA will move the system
from the lower left to upper right in Fig.  \ref{BBAR_IDIOT}$B$, eventually reaching a
steady-state current $J=p/4$.
  
Despite the apparent fundamental importance of the kinetic binding, or ``on'' rate in translation,
there are no systematic and independent measurements of $k$ in the literature.  The required
independent estimates of $k$ may be achieved by perhaps combined kinetic and affinity measurements of
the association of a minimal set of components, including only the ribosomes and a portion of the 5'
initiation codons and cofactors. For the off rate $\beta$, similar ideas can be employed. The
tRNA or ribosome release factor concentrations for the last codon can also be adjusted to
tune the off rate $\b$.

\vspace{2mm}

\noindent {\bf Codon and UTR Length Dependences} 
\vspace{1mm}

\noindent In experiments where it is possible to vary the number of codons $N$, the typical
harmonic distance $R$ can also be tuned. The phase diagrams in Figs.  \ref{PHASE0}, \ref{PHASE1},
and \ref{BBAR_IDIOT} all correspond to different regimes of Eq. \ref{JEXACT} in the large $N$
limit. In practice Eq. \ref{JEXACT}, is no longer sensitive to $N$ for $N\gtrsim 10$; however,
the harmonic distance $R$ between initiation and termination sites continues to increase as
$\sqrt{N}$, affecting the local concentration $C(a)$, and thus the effective {\it parameter} $\a
= kC(a)$ in Eq.  \ref{JEXACT}. As shown in Fig.  \ref{PHASE1}$C$, increasing $R/a$ shifts the
phase boundaries to the right, making the maximal current phase (III) harder to attain unless $k$
or $C_{\infty}$ is concomitantly increased. However, due to the $\sqrt{N}$ dependence, this
effect would be relatively weak for all but enormous values of $N$. Hence we have chosen the
qualitatively reasonable value $R/a=10$ in Figs.  \ref{BBAR_IDIOT}$A,B$.  

Although there may be a weak increase in $R/a$ as one increases the mRNA length, the effects of
increasing the coded sections ($N$) or the noncoded sections (the untranslated regions $m$,$n$),
can be different depending on $U_{0}$.  For large $U_{0}$, looped configurations dominate and the
distance between initiation and termination sites will be more sensitive to $m+n$, the shortest
distance between them (cf. Fig. \ref{fig2}$B$). The effect of lengthening $m+n$ on $R/a$ in the
high $U_{0}$ regime is clearly shown in Fig. \ref{REFF}$A$.  For small $U_{0}$, open
configurations dominate and the short segments $m$ and $n$ at the two ends do very little to
affect $R/a$ relative to $N$. Thus, although length dependences are expected to be weak,
increasing the codon length $N$ would more likely increase $R/a$ (and hence decrease throughput
$J$) in the small $U_{0}$, or repulsive limit. Conversely, increasing $m$,$n$ would more likely
increase $R/a$ when $U_{0}$ is large and loops dominate the mRNA conformations.

\vspace{2mm}

\noindent {\bf Initiation-Termination Cooperative Effects} 
\vspace{1mm}

\begin{figure}[!h]
\begin{center}
\includegraphics*[width=3.3in]{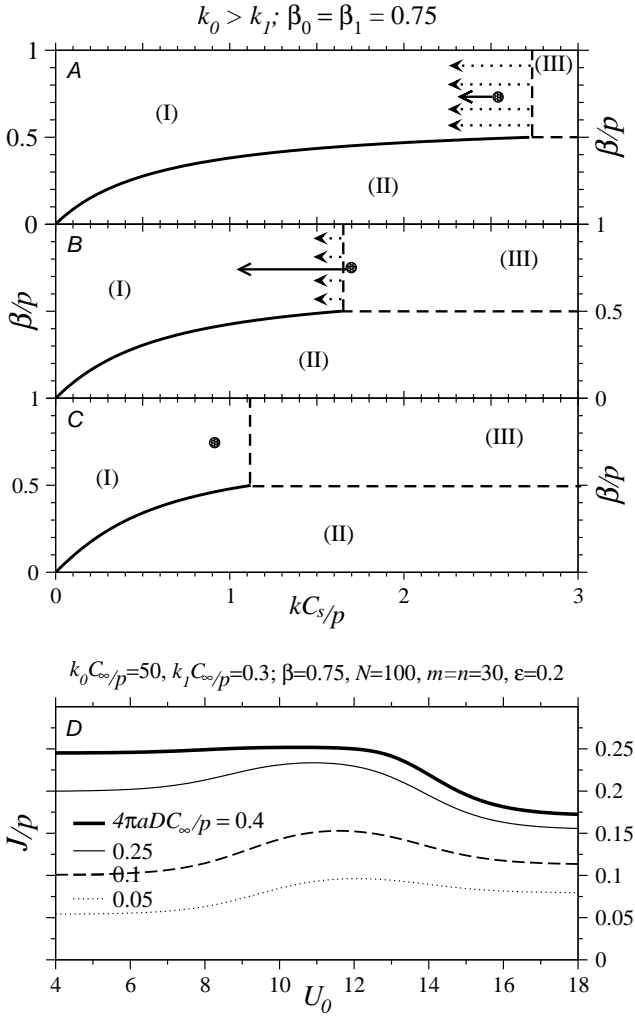}
\end{center}
\caption{The current (Eq. \ref{JEXACT}) as a function of
$U_{0}$ when the ribosome ``on'' rate $k$ can be modified by direct
interactions with elongation factor and PAB proteins. The Gaussian chain
approximation is used with persistence length $\ell = a$. $(A-C)$ show
hypothetical, qualitative  trajectories in the presence of a changing
phase diagram. As $U_{0}$ is increased, $R$ decreases. With $4\pi a D
C_{\infty}/p = 0.6$ fixed, the phase boundaries shown in $A-C$
correspond to $R/a = 25, 3, 3/2$, respectively. In addition, if $k_{0} >
k_{1}$, the effective binding rate $k_{\it eff}C_{\infty}/p$ also
decreases with increasing $U_{0}$, resulting in the trajectories
indicated by the dot.  $(D)$ Currents for $k_{0}C_{\infty}/p=50$ and
$k_{1}C_{\infty}/p=0.3$ and $N=100$.  The weak maximum appears only for
small $4\pi aDC_{\infty}/p$.}
\label{CONTROL}
\end{figure}

\noindent We have so far considered only the effects of the binding energy $U_{0}$ on loop
formation, $1/R$, and the resulting local ribosome concentration at the initiation site. 
However, evidence suggest that contact between elongation factor proteins and/or poly(A) tail
proteins can enhance or suppress the kinetic binding rates $k$ through direct molecular contact
and cooperativity \cite{SCANNING,MUNROE,SAC97,SAC00}. There is the possibility that in looped
states, PAB's can interact with {\it initiation} machinery and modify $k$, and/or elongation
factors can assist or hinder detachment of ribosomes at {\it termination}.  Modification of $k$
and/or $\beta$ through direct contact between proteins associated near the initiation and
termination sites may be an additional mechanism by which translation rates can span the regimes
shown in Figs. \ref{PHASE1}$B,C$ and \ref{BBAR_IDIOT}. Qualitatively, the experimental finding
that contact between the mRNA ends affect the initiation or possibly termination processes can be
modeled by assuming {\it effective} ``on'' or ``off'' rates


\begin{equation}
\begin{array}{cc}
k_{\it eff}[U_{0}] = k_{0}(1-P_{\it loop}) + k_{1}P_{\it loop} \\[13pt]
\beta_{\it eff}[U_{0}] = \beta_{0}(1-P_{\it loop}) + \beta_{1}P_{\it loop},
\label{KEFF}
\end{array}
\end{equation}

\noindent where $k_{0},\beta_{0}$ and $k_{1},\beta_{1}$ are the binding and ``off'' rates when the mRNA is open
and looped, respectively.  As $U_{0}$ is varied, both the intrinsic rates as well as the sink-source separation
$R$ are modified. Using expression \ref{KEFF} for $k$ and $\b$ in equations \ref{KSTAR} and \ref{BSTAR}, the
dependence of $J$ on the binding energy $U_{0}$ can be mapped.  A number of qualitatively different scenarios
are possible. If $\beta_{0}=\beta_{1}$ but $k_{1}>k_{0}$, the current is a monotonically increasing function of
$U_{0}$ because the binding rate increases and the ribosome source (3' terminus) is brought closer.  Both of
these effects monotonically increase the steady-state current.  However, if for fixed $\beta$, $k_{1} < k_{0}$,
then these two effects can partially balance each other and there is the possibility of a maximum in
$J(U_{0})$. A maximum occurs when initially, as $U_{0}$ is increased, the decrement in $k_{\it eff}$ cannot
keep up with the enhancement in local ribosome concentration due to the increasing likelihood of loop formation
({\it i.e.} the shifting of the high current phase boundary to lower $k_{\it eff}$). However, if $k_{1}$ is
sufficiently small, $k_{\it eff}$ eventually diminishes, such that one arrives at the low density, low current
regime.  These effects are illustrated in the sequence of figures \ref{CONTROL}$A-C$. The steady-state current,
self-consistently calculated from Eqs.  \ref{JEXACT}, \ref{KC}, and \ref{KEFF}, has a possible maximum and is
shown as a function of $U_{0}$ in figure \ref{CONTROL}$D$. Here, we have chosen $k_{0}C_{\infty}/p=50,
k_{1}C_{\infty}/p=0.3, \beta=0.75, N=100, m=m=30, \varepsilon=0.2, a=1$, and $\delta=0.1$. Only certain sets of
parameters permit a maximum. Small values of $4\pi aDC_{\infty}/p$ and large $N$ result in the largest
maxima. For large values of $4\pi aDC_{\infty}/p$, diffusion is fast, local ribosome concentrations are not
significantly depleted by the initiation site, and the high current regime is already pushed to low values of
$kC_{\infty}/p$. Therefore, increasing $U_{0}$ and decreasing $R$ does not further drive the high current
regime towards significantly lower $kC_{\infty}/p$.  For essentially the same reason, smaller $N$ enhance
ribosome recycling, increasing the current at low $U_{0}$,  thereby rendering the maximum in $J$ to lower
values of $U_{0}$. As illustrated in the exampled given in figure \ref{CONTROL}$D$, increases of $\sim 50-60\%$
above the background currentare possible as $U_{0}$ is varied. Thus, we see that the two processes, direct
molecular catalysis of initiation and termination, and ribosome diffusional depletion, balance each other and
may provide delicate control mechanisms during later stages of gene regulation.

\vspace{2mm}

\noindent {\bf Kinetic {\it vs.} Thermodynamic Control} 
\vspace{1mm}

\noindent Finally, we point out that our analysis has been confined to the steady-state (for the
bulk ribosome diffusion and individual ribosome movement along the mRNA) and thermodynamic
equilibrium (for the statistics of the polymer statistics).  Since it is possible for diffusion
and ribosome elongation along the mRNA to reach steady-state before the mRNA chain reaches
conformational equilibrium (in the presence of loop-forming proteins), a possibility exists for
``kinetic {\it versus} thermodynamic control'' for the measured ribosome throughput.  Although
the loop-binding energy $U_{0}$ determines the equilibrium distribution of open and closed mRNA
conformations via $P_{loop}$, the kinetics of loop opening and closing are determined by energy
{\it activation} barriers of the loop binding proteins.  For example, if the activation energy
for creating a looped state is high, the mRNA may sample only unlooped conformations on time
scales of the ``steady-state'' (with respect to the TASEP and diffusion).  In this scenario, the
effect of the loop binding protein does not arise and the harmonic distance $\langle R\rangle$
would appear to be that associated with an open chain ($U_{0} \rightarrow -\infty$ in Fig. 
\ref{REFF}$A,B$).  Conversely, if the the mRNA chain happens to be in a looped conformation and
the free energy barrier for dissociation of the loop is large, the measured current may be that
corresponding to only a closed mRNA loop (mimicking the case $U_{0}\rightarrow \infty$).  This is
likely to occur if the measurement time $\tau \ll \tau_{diss} \sim De^{-U^{*}}$, where
$\tau_{diss}$ is the spontaneous dissociation time (or the Kramers escape time) and $U^{*}$ is
the activation barrier energy/$(k_{B}T)$.  The activation energy $U^{*}$ depends on the specific
molecular details of the loop-forming proteins; however, measurements using fluorescence
quenching can be used to independently determine the distribution of times the mRNA chain is
looped or unlooped \cite{LIBCHABER}. Only when $U_{0}$ or $U^{*}$ are large does ribosome
recycling get significantly enhanced by loop formation. Transient measurements, as well as
fluctuations of the measured throughput, is beyond the scope of the paper.

\vspace{4mm}
\noindent {\bf SUMMARY}
\vspace{2mm}

We have constructed a simple model and road map for the possible physical
effects at play during translation.  The model incorporates driven diffusive
motion which obeys exclusion statistics for ribosomes along mRNA. The initiation
and termination sites are considered as sinks and sources of ribosome
concentration, described by the steady-state diffusion equation (Laplace's
equation).  The average conformations of the mRNA chain define the typical
initiation-termination distance which determines the how terminated ribosomes
directly diffuse back to the initiation site and affect the local concentration
there. This local concentration is a parameter (the injection rate) in the
exclusion process, but also depends on the overall ribosome throughput (the
strength of the sink and source). Thus, the current  $J$ needs to be solved
self-consistently.  Direct cooperative enhancement of kinetic binding and ``off''
rates were also incorporated.  Although it is thought that the rate-limiting step
is binding and initiation of ribosomes at the initiation site
\cite{CLEMENS,SHIT}, the fact that polysomes have been found to exist in both
high and low ribosome occupancy states suggests that under physiological
conditions, steady-state ribosome fluxes can span the regimes defined by the phase
diagrams depicted Figs.  \ref{PHASE0} and \ref{PHASE1}$B,C$.  At high occupancy, the rate
limiting step is the off rate $\beta$ which controls the steady-state flux (cf.
Phase (II) in Fig.  \ref{PHASE0}).  Ribosome depletion by the sink and
replenishment by the source can drastically affect the constant $k,\beta$ phase
diagram, as shown in Figures \ref{PHASE1}. The critical values of $k^{*}C_{\infty}/p$
that define the the left boundary of the maximal current phase (in the
$N\rightarrow \infty$ limit) is most sensitive to the dimensionless parameter
$4\pi aDC_{\infty}/p$ when $4\pi aDC_{\infty}/p\simeq 0.15-0.3$. For sufficiently small
$4\pi aDC_{\infty}/p$, the effective injection rate cannot reach $1/2$ and the maximal
current phase cannot be attained.  When $N\neq \infty$, the explicit currents
were computed from Eq. \ref{JEXACT} and plotted in Figure \ref{JEXACTFIG}.  Given
the possibility of cooperative interactions in looped mRNA configurations, we have
also found a maximum in ribosome throughput as a function of loop-binding energy
$U_{0}$.

Many molecular and chemical details have been neglected.  As mentioned, we have ignored the
fact that numerous components must assemble before initiation and have modelled only an
``effective'' rate-limiting component.  The surface concentration parameter $C(a)$ in our
model would be an effective concentration reflecting the local density of ribosomes capable of
initiation.  Proposed mechanisms of ribosome scanning \cite{SCANNING}, whereby ribosomes
attach to segments of mRNA and undergo one-dimensional diffusion before encountering the
initiation site, can be adequately modeled with the present approach if one assumes that the
rate-limiting step is initial adsorption onto an mRNA segment.  Furthermore, we have assumed
that the ribosomes do not detach from the mRNA until they reach the termination site and that
their forward hopping rates are uniform across the whole coding region. Finally, in our simple
polymer model, we have neglected both self-avoidance (of both chain-chain and chain-ribosome
exclusion) and the fact that the effective persistence length may varying along the mRNA,
depending on the local ribosome density.  

Despite these simplifying assumptions, we find that {\it qualitatively}, subtle control
mechanisms can come into play, depending on biologically reasonable physical parameters. 
Although there are numerous experiments probing translation, both {\it in vivo} and {\it in
vitro}, many different systems and physical conditions are employed, rendering quantitative
comparison with measurements difficult. Nonetheless, our model suggests new measurements
that can be used to qualitatively probe the various physical hypotheses and exhibit our
predicted physical trends. For example, the effective $C_{\infty}$ can be varied in a number
of ways to test with the predicted current regimes.  Occupancy along the mRNA can also be
correlated with the high, low, and intermediate density phases.  Additionally, the noncoded
regions between the elongation factors and the initiation site, and the termination site and
the poly(A) tail-bound PAB can be varied to test possible cooperative interactions defined by
Eq.  \ref{KEFF}.  Since the loop formation probability $P_{\it loop}$ depends on the total
statistical length $L_{T}$, which is dominated by the length of the coding region ($L_{N}a^{2}
\gg (m+n)\e^{2}$), varying $m$ and $n$ would affect, through the likelihood of molecular
contact in the looped states, only $k_{\it eff}$ and $\beta_{\it eff}$, respectively.  The
actual probability of loop formation $P_{\it loop}$, and hence $R$,  would not be
significantly affected.  Chemical modification of the elongation factors or the PAB's {\it
would} affect  $U_{0}$, and hence $k_{\it eff},\beta_{\it eff}$, and $R$ through $P_{\it
loop}$. Using micromanipulation techniques \cite{BUSTA}, it might also be possible to fix the
initiation-termination distance {\it in vitro}.

Numerous extensions to the presented models can be straightforwardly incorporated to more precisely
model the chemical and microphysical processes.  Codon and tRNA concentration-dependent variations in
the internal transition rates $p$ \cite{CODON}, as well as random detachment processes, can be
implemented using simple lattice simulations.  Sites along the mRNA chain at which ribosomes pause can
be treated as ``defects'' in a TASEP and the whole process can be treated with mean-field theory
\cite{KOLO}.  Multiple coding regions in prokaryotic translation (Shine-Dalgarno sequences) can be
modeled as a sequence of initiation (sinks) and termination (sources) sites. Similarly, cap-independent
initiation at internal ribosome entry sites \cite{SCANNING,MAR01,SAC97} (IRES) can also be treated as
sinks within our basic model. Translation of ER-associated mRNA further involve ribosomes that attach
the mRNA at certain points on the ER membrane.  In this case, one expects that density of cytoplasmic
and ER-bound ribosomes to have a strong effect on localization of mRNA to ER and overall translation
rates.  One can also consider cases where the protein product itself is a ribosome product necessary
for its self-translation; this processes would result in initially autocatalytic protein production. 
Although these more complicated and interesting extensions have not been considered here,  the simple
models we have presented represent a first step towards the rich problem of identifying and quantifying
the physical and biological mechanisms that control late stages of expression.

\vspace{2mm}

The author thanks S. Bump, T. Chang,  D. Herschlag, G. Lakatos, E. Landaw, J. 
Rudnick, and M. Suchard for vital discussions and helpful suggestions. The
author especially wishes to thank an anonymous reviewer for correcting an error
in the original formulation of the adsorption  boundary conditions. This work
was supported by the National Science Foundation through grants DMS-9804370 and
DMS-0206733.  

\begin{appendix}

\section{PHYSICAL ASSUMPTIONS AND MATHEMATICAL APROXIMATIONS}

Although our model arrives at a number of conclusions that are developed by
combining three different physical theories, the assumptions and approximations
used in each are well developed in 
the condensed matter physics and biophysics literature. Here, we summarize the main
physical assumptions and review the mathematical approximations used:

\vspace{2mm}

\noindent $\bullet$ {\it Steady-state and equilibrium assumptions}: Ribosome diffusion
and motion along the mRNA are treated within steady state, while the
configurational distribution of the mRNA polymer is not directly coupled to
ribosome diffusion or motion, and is considered in thermodynamic equilibrium. 
The inverse ``harmonic distance'' $1/R$ is determined from {\it equilibrium}
mRNA configurational distributions, but parametrically influence the
nonequilibrium steady-state processes of diffusion and the TASEP. Equilibration
times of unentangled polymers and diffusion times over the length of the mRNA
are on the order of milliseconds to seconds, while the relaxation to steady-states
in the TASEP occur over seconds to on the order of a couple minutes. Thus, on
experimental time scales longer than these, transients in the ribosome throughput
have dissipated, and the steady-state and equilibrium assumptions are
appropriate.

One might be tempted to formulate the specific mechanisms in terms of the common
notions of reactions being kinetically or thermodynamically controlled.  In
this biochemical terminology, the TASEP is kinetically controlled, since the ribosomes take
irreversible steps as each amino acid is added during elongation. The mRNA
configurations, computed under equilibrium conditions, are by definition
thermodynamically controlled. However, each of the proposed mechanisms is a simple,
single, independent  process, the notion of kinetic control {\it versus}
thermodynamic control is irrelevant. Within each mechanism, there are no alternate
``reaction paths'' or outcomes for kinetic or thermodynamic control to apply.
However, it is possible that the mRNA conformations and the
binding protein-mediated loop formation does not reach equilibrium on the
time scale of measurements of ribosome throughput. This possibility is also
discussed in the Experimental Consequences and Proposed Measurements 
section.

\vspace{2mm}

\noindent $\bullet$ {\it Gaussian chain polymer model for mRNA}: Unlike tRNA, the coding
regions of mRNA is relatively devoid of secondary structure. The single-stranded
mRNA is treated using standard statistical physics of polymers that assumes
nonintersecting random walks of step size defined by the polymer persistence length.
For single-stranded mRNA without adsorbed proteins, the persistence length $\approx
2-3$ bases. When loaded with large ribosomes, we assume that the persistence
length is on the order of the ribosome size and that it is approximately uniform
along the chain. Although the ribosome loading might varying slightly along the
chain, this variation occurs only near the ends and does not appreciably affect the
equilibrium end-to-end distributions. Although we treat only phantom
(nonintersecting) polymers, effects due to the binding of finite-sized PABs and cap
proteins are explicitly treated when computing the end-to-end distribution functions
in the small distance regime where steric exclusion of the end proteins are
important.

\vspace{2mm}

\noindent $\bullet$ {\it Single component ``ribosomes''}: The assembly of ribosomes
before or during adsorption onto the initiation site can be modeled as an effectively
single, rate-limiting component that undergoes standard diffusion in the bulk
solution. Including more chemical details will not qualitatively alter our results,
since in diffusive steady-state, all species' concentrations  would be spatially
distributed as $1/R$ and parametrically affect the TASEP in the same qualitative
manner.

\vspace{2mm}

\noindent $\bullet$ {\it Equal particle and step sizes}: Ribosomes
moving along mRNA are treated with a discrete TASEP where the 
step size is exactly equal to the particle diameter. However, ribosomes 
are large and occlude $\sim 10$ codons so that they move one particle diameter
only after about $q=10$ steps (amino acid transfers). Nonetheless, the 
qualitative behavior of the currents for different $q$ remain unchanged.
For the sake of simplicity and clear analytic expressions
(Eqs. \ref{KSTAR}, \ref{BSTAR}, and \ref{JEXACTB}), we have restricted our 
analysis to $q=1$. Exact large $N$ asymptotic expressions for the 
steady state current for general $q$ are given in Appendix D.

\vspace{2mm}

\noindent $\bullet$ {\it Uniform elongation step rates in the TASEP}: The analytic
solutions represented by Eqs.  \ref{JEXACT}, \ref{SEXACT}, and \ref{ASEMEQN} are
based on uniform elongation rates $p$ along the mRNA. It is known that $p$ can vary
by factors of 2-10 \cite{CODON}, depending on the codon in question and the
availability of the associated tRNA. As a first step, we have simply assumed a
scenario in which the elongation rates do not vary appreciably over the coding
region. More elaborate models that include specified elongation rates $p_{i}$ across
the mRNA chain would require extensive simulations for each realization of
$\{p_{i}\}$.

\vspace{2mm}

\noindent $\bullet$ {\it Bulk diffusion limited adsorption}: Ribosomes, or the relevant
rate-limiting component of a ribosome, diffuses in bulk and directly attaches to the
initiation site. Capture of the ribosome by the initiation end of the mRNA may occur
in a two-step process of nonspecific adsorption from bulk, followed by linear
diffusion along a segment of the mRNA, before ultimately interacting specifically
with the initiation site \cite{VONHIPPEL,STANFORD}. Although studied in the context of linear diffusion and
search along DNA \cite{BERG}, direct evidence for such scanning mechanisms in the
initiation of mRNA translation has been hard to obtain \cite{SCANNING}.
For example, secondary structure in the form of small mRNA knots near the 5'
region must be melted before efficient ribosome scanning can occur \cite{KOZAK}.
Nevertheless, one-dimensional diffusion of ribosomes along the
mRNA near the initiation sight {\it is} implicitly included in our model.
The conjectured scanning mechanisms suggest that ribosomes 
scan locally near the start codon \cite{SCANNING,MILLER97}. Thus, if ribosome 
recycling via diffusion through the bulk is rate-limiting, the scanning region
near the initiation where the linear diffusion occurs can be considered  
as binding region of larger {\it effective} capture radius $a$.

\section{MEAN FIELD ANALYSIS FOR LARGE PARTICLES}

Consider identical particles that are driven through a long one-dimensional lattice of $L$ sites. The
lattice is discretized into steps of unit length (a step size corresponding to a codon step), while the
particles are of integer size $q\geq 1$.  For each particle to move a distance roughly equal to its
diameter, $q$ consecutive steps must be taken. Thus, we expect that effectively, the mean current
would be approximately described by equations \ref{JEXACT} or \ref{ASEMEQN} but with $p$ replaced by
$p/q$. A mean field model for the asymmetric exclusion process containing particles that occupy $q$
substrate lattice sites (mRNA codons) has been solved.  The analysis is beyond the scope of this paper,
but the resulting steady-state currents follow the same qualitative ``phase diagram" (Fig.
\ref{PHASE0}) as the TASEP with particles of size $q=1$. That is, for large entrance and exit rates,
there is a maximal current phase (III), bounded by low (I) and high (II) density phases. The effects of
increasing the particle size to $q>1$ only quantitatively changes the values of the currents in each of
these phases, and can be straightforwardly integrated into the present study.  

The general (for all particle sizes $q$) result for the steady-state currents 
in the infinite chain length limit are

\begin{equation}
\begin{array}{rll}

\mbox{(I)} & \displaystyle  \a<{p\over 2}, \a<\b & \:\, \displaystyle J \equiv J_{L} ={\a(1-\a/p)\over 
1+(q-1)\a/p} \\[13pt]
\mbox{(II)} & \displaystyle  \b< {p\over 2}, \b<\a & \:\, \displaystyle J \equiv J_{R} =  {\b(1-\b/p)\over 
1+(q-1)\b/p} \\[13pt]
\mbox{(III)} & \displaystyle  \a,\b\geq {p\over 2} &\:\, \displaystyle J\equiv J_{max}=
{p\over (\sqrt{q}+1)^{2}}. 
\end{array}
\label{ASEMEQNQ}
\end{equation} 

\noindent These results have been verified to be exact (to within numerical precision) by
extensive Monte-Carlo simulations.  Note that for large $q$, the maximal current $J_{max}$
is that given by Eq. \ref{ASEMEQN} but with $p \rightarrow p/q$. These results only serve
to quantitatively shift the phase boundaries between the different current regimes and
decrease the magnitude of the currents. For example, if $q=2,3$, the phase boundary
between the low density and the maximal current regime occurs at $\a/p = 0.41,0.37$,
respectively, rather than at 0.5.  For the sake of simplicity and manageable algebraic
expressions, we have in this study only considered the $q=1$ case. Our analysis should be
applied to the mRNA translation problem with the understanding that $p$ in Eq.
\ref{ASEMEQN} and subsequent equations is roughly the rate for a ribosome to move its molecular
size, not the rate for an individual tRNA transfer. If, however, the above expressions were used, 
then $p$ in expressions \ref{ASEMEQNQ} would be identified with the typical single amino acid 
transfer rate.

\section{OPEN CHAIN PROBABILITY DISTRIBUTIONS}

\begin{figure}
\begin{center}
\includegraphics[height=2.6in]{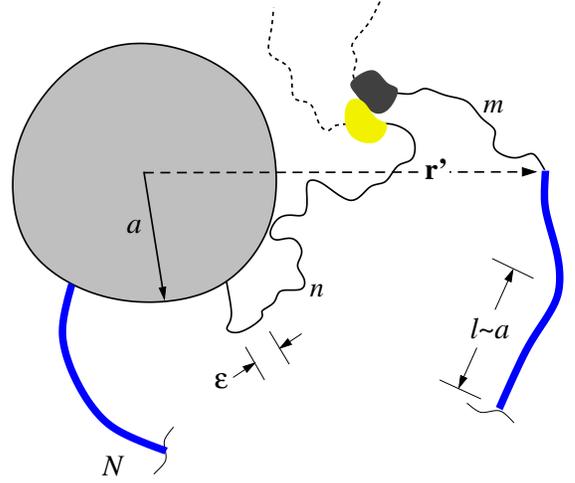}
\end{center}
\caption{Schematic of the geometry near the
initiation-termination end of a looped mRNA. The mRNA loop binding
factors are shown in yellow and black, while a ribosome of radius $a$ is
situated at the initiation site (not drawn to scale). $m$ and $n$
correspond to the number of bases of the UTR's which are assumed to be
relatively protein-free and have short persistence length $\varepsilon$.
Here, the persistence lengths in the coding regions (thick curve,
described by the TASEP) is $\ell \sim a$.}
\label{fig3}
\end{figure}

Consider the probability distribution $W(\r\vert\mbox{open})$ of the
initiation-termination separation {\it in the absence} of loop
formation.  Since the ribosome can be much larger than the typical
persistence length in the noncoding region of single-stranded mRNA, $a
\gg \e$.  For $a \sim 10\e$, $a \not\ll L_{mn}$, unless the noncoding
regions are very long, with $m+n \gg 100$.  For shorter noncoding
regions, the expression for $W(\r;L_{mn}|\mbox{open})$ must be evaluated
more carefully, particularly for small $r$, in order to compute $\int
\mbox{d}\r W(\r)/r$ correctly.  Assume the termination site starts a
random walk from any position on the sphere.  Details of the different
segments of mRNA are shown in figure \ref{fig3}.  The problem maps to
that of heat diffusion from a sphere of size $a$ with reflecting
boundary conditions and an instantaneous uniform temperature source on
the surface. The probability that the initiation site (that is linked to
the termination site via $m+n$ persistence lengths) is within $r$ of the
sphere can also be described by the temperature near a sphere with an
exterior instantaneous source of temperature. The diffusion equation for
the probability distribution $W(\r;L_{mn}|\mbox{open}) \equiv W$ obeys

\begin{equation}
\dot{W}(\r,t) = \kappa \Delta W(\r,t)
\label{HEAT}
\end{equation}

\noindent where the thermal conductivity is associated with the squared persistence
length, $\kappa \leftrightarrow \e^{2}/6$, and time corresponds to the length $t
\leftrightarrow m+n$.  The initial and boundary conditions corresponding to a chain
that originates from the surface of the otherwise impenetrable ribosome particle are

\begin{equation}
\partial_{r}W(r=a) = 0, \,\, W(\r,t=0) = {\delta(r-a)\over 4\pi a^{2}},
\end{equation}
 
\noindent where we have assumed spherical symmetry. 
Following Carslaw and Jaeger (1959), we define $W(r,t) = f(r,t)/r$ to reduce
(\ref{HEAT}) to $\dot{f}(r,t) = \partial_{r}^{2}f(r,t)$, with boundary conditions

\begin{equation}
\partial_{r}f(r=a) = {1\over a} f(a),\,\, f(r,t=0)={r\delta(r-a) \over 4\pi a^{2}}.
\end{equation}

The solution for $f(r,t)$ is found using Laplace transforms, and is

\begin{equation}
\begin{array}{l}
\displaystyle f(r,t) = {1\over 8\pi a \sqrt{\pi \kappa t}}
e^{-(r-a)^2/(4\kappa t)}- \\[13pt]
\displaystyle \hspace{1cm}
{e^{r/a-1}e^{\kappa t/a^{2}}\over 4\pi a^{2}}
\mbox{Erfc}\left[{r-a \over 2\sqrt{\kappa t}}
+{\sqrt{\kappa t}\over a}\right].
\end{array}
\end{equation}

\noindent The probability distribution is thus

\begin{equation}
\begin{array}{l}
\displaystyle W(\r,L|\mbox{open}) = {\sqrt{3}e^{-3(r-a)^2/(2N\ell^{2})} \over
2(2\pi)^{3/2}\sqrt{N}a\ell r } - {e^{r/a-1}\over 4\pi a^{2}r}\\[13pt]
\:\hspace{1cm} \displaystyle e^{N\ell^{2}/(6a^{2})}
\mbox{Erfc}\left[\sqrt{{3\over 2N}}{r-a \over\ell}+\sqrt{{N\over 6}}{\ell\over a}\right].
\end{array}
\label{Wmn}
\end{equation}

\noindent Note that {\it if} $a/L \ll 1$, as it is for
$L = L_{N}$, equation (\ref{Wmn}) would be approximately

\begin{equation}
\begin{array}{ll}
\displaystyle W(\r,L|\mbox{open}) \sim & \displaystyle \left({3 \over 2\pi L^{2}}\right)^{3/2}
e^{-3r^{2}/(2L^{2})}\times \\[13pt]
\: & \displaystyle \bigg[1-36{a^{2}\over L^{2}}
\left(1-{x^{2}\over L^{2}}\right)+O\left({a^4\over L^{4}}\right)\bigg],
\end{array}
\end{equation}

\noindent which reduces to end-to-end probability distribution for a Gaussian random
chain. However, since $a/L_{mn} \not\ll 1$, we need to use the full expression Eq. 
\ref{Wmn} for the loop contribution (Eq. \ref{WLOOP}) in the  calculation of $W_{\it
eff}(r)$ and $1/R$.

For the WLC, an approximate probability distribution function can be reconstructed from commonly
used phenomenological force-extension relationships. If the force-extension interpolation given by
Marko and Siggia \cite{SIGGIA} is shifted to take into account the finite-sized origin,

\begin{equation}
f(z) = \ell^{-1}\left[{1\over 4\left(1-{z-a\over N\ell}\right)^{2}}+{(z-a)\over N\ell} 
-{1\over 4}\right].
\end{equation} 

\noindent The initiation-termination distance distributions can be 
estimated using

\begin{equation}
W_{WLC}(\mbox{open}\vert \r) \approx {\exp\left[-\int_{N\ell}^{r} f(z) dz\right]
\over \int_{a}^{N\ell}\mbox{d}r \exp\left[-\int_{N\ell}^{r} f(z) dz\right]}.
\end{equation}

\begin{figure}[!h]
\begin{center}
\includegraphics[height=5.1in]{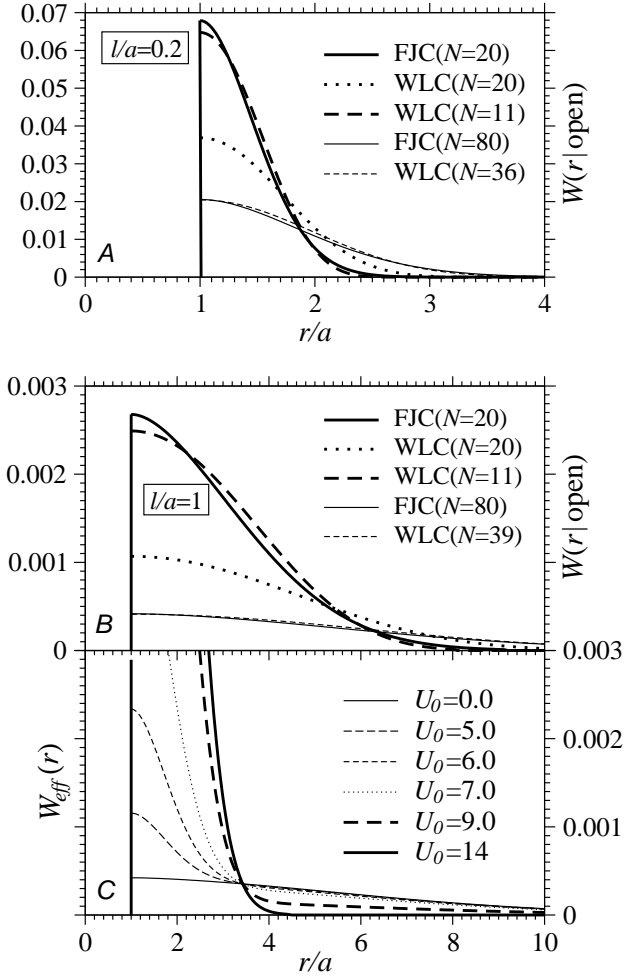}
\end{center}
\caption{$(A)$ FJC and WLC models for $W(\r\vert\mbox{open})$
for $\ell/a = 0.2$.  The WLC distribution approximates that of the FJC
if the effective number of persistence lengths $N$ is reduced. This
reduction compensates for the stiffness of the chain that tends to give
more weight at larger distances.  $(B)$ FJC and WLC distributions for
$\ell/a=1$. Note the heuristic cutoff applied to the WLC model at $r =
a$. As expected, for equal $N$, the WLC model gives a typically larger
separation and hence smaller $a/R$; however, $a/R \propto N^{-1/2}$ for
$N\rightarrow\infty$ in all cases.  $(C)$ The effective end-to-end
distance distribution $W_{\it eff}$ constructed from
$W(r\vert\mbox{open})$ via equations \ref{PLOOP0} and \ref{WLOOP}.}
\label{PROB}
\end{figure}

\noindent This end-to-end probability distribution from both FJC and WLC models are plotted in
Figs. \ref{PROB}$A,B$.  The WLC model gives qualitatively similar distributions to
those of the FJC model, provided the contour length is appropriately reduced. Furthermore, the
WLC and FJC models provide qualitatively similar averages $\langle a/r\rangle$ if the $N$ used
in the WLC is sufficiently reduced. Upon using Eqs. \ref{WLOOP} and \ref{WEFF}, one can
compute the effective end-to-end distribution of a chain with segments of different
persistence length and with attached loop binding proteins, as shown in Fig. \ref{PROB}$C$.

\section{ASYMPTOTICS FOR $J_{N}$}

Asymptotic expressions for the steady-state current given by Derrida {\it et al.} \cite{DER93} are
valid only far from the phase boundaries.  However, in our present model, we are interested in how a
change in the mRNA length $N$ allows the system to cross over from one behavior to another.  For the sake
of completeness, we derive limiting forms for the current $J_{N}$ near phase boundaries.  An asymptotic
expansion in the rates about $\a=1/2$ is taken first, with $N$ fixed.  From the exact expression Eq.
\ref{SEXACT} given by \cite{DER93}, we find the following asymptotic expansion

\begin{equation}
\begin{array}{ll}
\displaystyle S_{N}(x=2)&  = 4^{N}{2\over \sqrt{\pi}}{\Gamma(N+1/2)\over N\Gamma(N)}\\[13pt]
\: & \displaystyle \sim 4^{N}{2\over \sqrt{N\pi}}\left[1-{1\over 8N} 
+ {1\over 128 N^{2}} + O(N^{-3})\right]
\end{array}
\end{equation}

\noindent  For $\b > 1/2$, and $\a = 1/2 + \e$, we take the large $N$ limit, but with $\e\sqrt{N}\rightarrow
0$. The resulting current across the maximal current-low density phase boundary is

\begin{equation}
\begin{array}{l}
\displaystyle J \sim {1\over 4}\left[1 + {1\over N} + 
{\b(\b-1) \over (2\b-1)^{2}N^{2}} + O(N^{-3})\right] + \\[13pt]
\displaystyle \hspace{3mm}{3\sqrt{\pi}\over 32}\left[\sqrt{N} + {52\b^2-52\b +17 
\over 8(2\b-1)^{2}\sqrt{N}} + O(N^{-3/2})\right]\e + O(\e^{2}).
\end{array}
\end{equation}

\end{appendix}

%
%

%
%

%
%

\end{document}